\documentclass{emulateapj}

\submitted{Scheduled to appear in \apj\ July 10, 2008}

\newcommand{\um}{$\mu$m}
\newcommand{\acet}{${\rm C}_2{\rm H}_2$}
\newcommand{\ctemp}{$[6.4]-[9.3]$}
\newcommand{\smpy}{M$_{\sun}$~${\rm yr}^{-1}$}
\usepackage{lscape}

\shorttitle{Metallicity Effects on Carbon Stars}
\shortauthors{Leisenring et al.}

\begin{document}


\title{Effects of Metallicity on the Chemical Composition of
Carbon Stars}


\author{J. M. Leisenring\altaffilmark{1}, F. Kemper\altaffilmark{2,1}, G. C. Sloan \altaffilmark{3}}


\altaffiltext{1}{University of Virginia, Astronomy Department,
Charlottesville, VA 22904} \altaffiltext{2}{Jodrell Bank Centre for
Astrophysics, University of Manchester, Manchester, M13 9PL, UK}
\altaffiltext{3}{Cornell University, Astronomy Department, Ithaca,
NY 14853}


\begin{abstract}
We present \emph{Spitzer} IRS data on 19 asymptotic giant branch
(AGB) stars in the Large Magellanic Cloud, complementing existing
published data sets of carbon stars in both Magellanic Clouds and
the Milky Way, to investigate the effects of metallicity on dust and
molecular spectral features arising from the circumstellar envelope.
We find that the C$_2$H$_2$ P and R branches at 7.5 micron are
affected by dust dilution at higher mass-loss rates -- albeit to a
lesser extent for sources in the Magellanic Clouds, compared to the
Milky Way -- while the narrow 13.7 micron C$_2$H$_2$ Q branch only
shows the effect of dust dilution at low mass-loss rates. A strong
metallicity dependence is not observed for the Q branch. Independent
of metallicity, we also provide an explanation for the observed
shifts in the central wavelength of the SiC emission feature, as we
show that these are largely caused by molecular band absorption on
either side of the dust emission feature, dominating over shifts in
the central wavelength caused by self-absorption. We have devised a
method to study the dust condensation history in carbon-rich AGB
stars in different metallicity environments, by measuring the
strength of the 11.3~\um\ SiC and 30~\um\ MgS features with respect
to the continuum, as a function of mass-loss rate. With this method,
it is possible to distinguish in what order SiC and graphite
condense, which is believed to be sensitive to the metallicity,
prior to the eventual deposit of the MgS layer.
\end{abstract}





\section{Introduction}

Stars with a main-sequence mass of $\sim$1-8~M$_{\sun}$ undergo an
asymptotic giant branch (AGB) phase which plays an important role in
the recycling of stellar material into the ISM. During the AGB
phase, a significant fraction \citep[up to 90\%;][]{wei87} of the
stellar mass is lost in a gradual stellar wind. Initially, this wind
results in a mass-loss rate of 10$^{-7}$~\smpy\ and eventually
evolves towards a superwind phase with $\dot{M}\sim10^{-4}$~\smpy\
\citep{hab96}. Given the shape of the initial mass function, AGB
stars contribute a substantial amount of gas and dust to the ISM
(about half the total recycled gas material) driving further
galactic evolution \citep{mae92}.

The mass-loss mechanisms still remain poorly understood even though efforts at modeling
are decades old \citep{sal74}. Thermal pulses enrich the star's outer layers by extending
a convective zone downward to the He shell and allowing carbon to be mixed to the surface
\citep{sch65,her07}. Material then flows from the stellar surface due to an unknown
mass-loss mechanism. Once the gas has cooled enough to allow the formation of molecules
and condensation of dust grains, the impeding radiation pressure from the central source
on these grains is sufficient to accelerate the grains and drive a dust-driven wind by
grain-gas drag \citep{hoy62,fle95,hof95}. The complex interplay between the internal
thermal pulses, the dust-driven wind, and envelope pulsations provides the theoretical
framework for the observed light curves and mass-loss rates \citep{whi97}.

The efficiency of grain formation and rate of grain growth is thus expected to affect the
mass-loss history of AGB stars. We can constrain dust formation within the circumstellar
envelopes of AGB stars by analyzing the isotopic compositions of certain grains found in
primitive meteorites. Since these meteorite compositions are found to differ greatly from
solar compositions, they must have originated prior to the formation of our solar system
and are thus dubbed `presolar.' Detailed studies of presolar grain characteristics (such
as sizes, crystal structures, and composition) can be accomplished after extraction from
the meteorites \citep{ber97,nit03,cla04,zin06}. These dust grains are virtually unaltered
from their formation state in AGB star and thus provide constraints on the physical
characteristics of the dust we observe in the circumstellar shells of Galactic AGB stars.
Silcon Carbide (SiC) and carbon graphite dust grains account for the majority of
discovered presolar grains and can be attributed to formation in Galactic carbon stars
\citep{ber06}. Due to the aggressive and destructive nature of analyzing the meteorites,
all but the most refractory grains become dissolved. Only recently have presolar
silicates been discovered meteorites and interplanetary dust particles \citep{mes05}.

\begin{figure*}[htp]
\begin{center}
\includegraphics[angle=90]{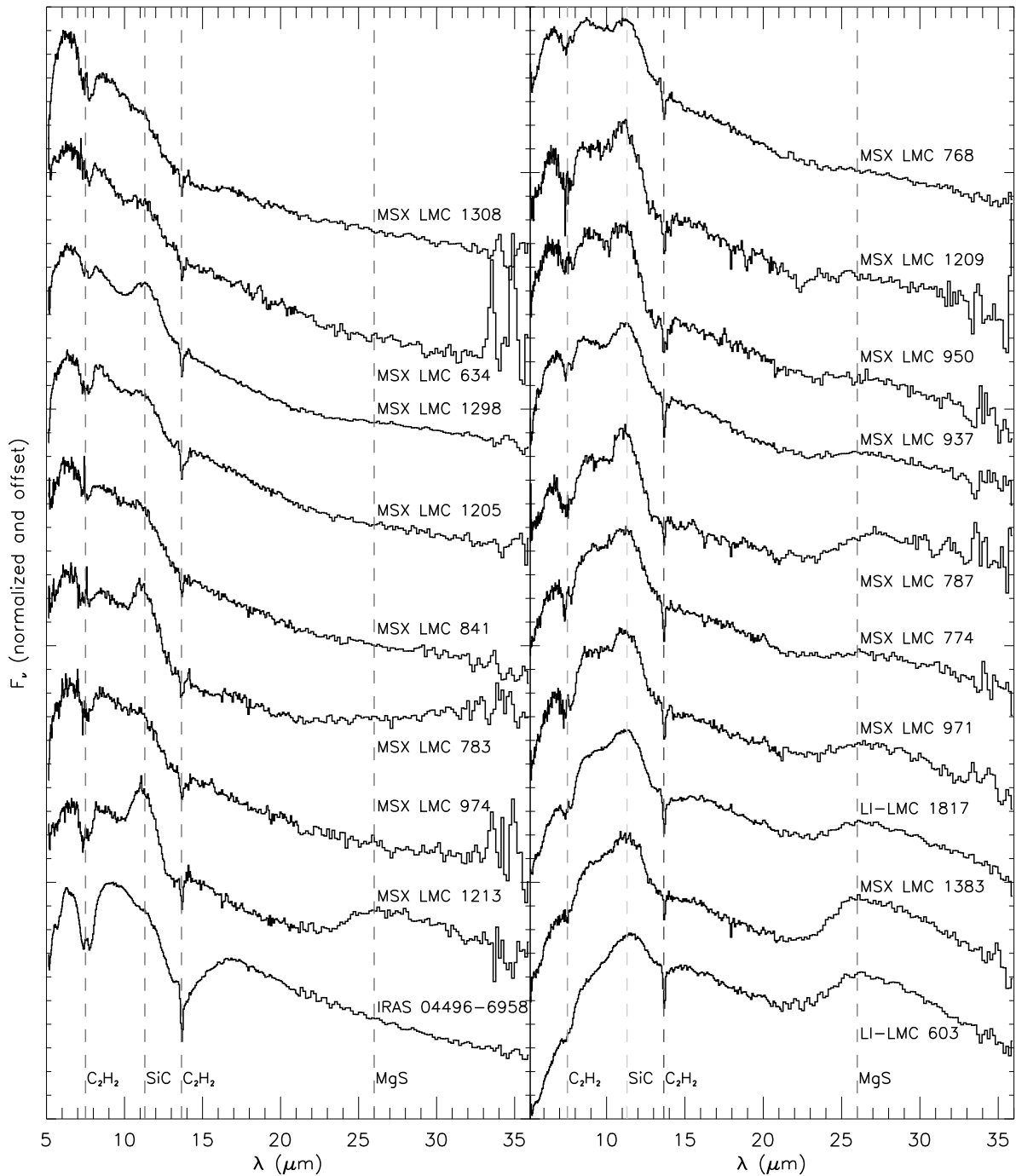}\\
\caption{{Our sample spectra of carbon-rich AGB stars observed in the LMC ranging from
5-36~\um. The spectra have been ordered by their \ctemp\ color from blue to red,
normalized by the maximum value of their fitted blackbodies, and offset for clarity. The
dashed lines indicate the positions of the \acet\ absorption features at 7.5 and
13.8~\um, the SiC emission features at 11.3~\um, and the MgS emission feature at
$\sim$26~\um.}} \label{fig:allplots}
\end{center}
\end{figure*}

Initially, the mass loss occurs from the outer layers of the star, and hence the chemical
composition of the recycled material reflects the original ISM composition from which the
star was formed. As mass loss progresses, dredge-up processes enrich the stellar ejecta
with nuclear fusion products (C, O), s-process elements (Ba, Pb), and lithium
\citep{dra03}. The nuclear fusion produces more carbon than oxygen \citep{lat00},
consequently causing the C/O ratio of the stellar atmosphere to increase from dredge-up
of material produced in the H- and He-burning layers. For a solar metallicity
environment, \cite{str97} finds that AGB stars above $\sim$1.5~M$_{\sun}$ will eventually
become carbon stars unless they exhibit the hot bottom burning phenomenon of more massive
stars \citep[$M\gtrsim5$~\smpy;][]{pol01}. However, those with a lower metallicity
initially contain less oxygen in their atmospheres than their metal-rich cousins and,
therefore, will more easily become carbon stars. Thus, as metallicity decreases, the
lower mass limit at which stars will evolve into carbon stars also decreases
\citep{ren81,vas93}. In addition, metallicity may have a profound effect on grain growth,
the number density of grains, and the grain size distribution, thus altering the
influence of the radiation pressure from the central star. The chemical composition of
material in the wind will also be affected by the metallicity.

\begin{deluxetable*}{lrrrrrrrcrr}
\tabletypesize{\scriptsize} 
\tablewidth{0pt} \tablecaption{Observed LMC Targets \label{tbl:obover}}
\tablehead{\colhead{Adopted} & \colhead{RA} & \colhead{Dec} &
\multicolumn{3}{c}{2MASS\tablenotemark{a}} & \colhead{MSX\tablenotemark{b}} &
\colhead{AOR} &
\colhead{Observation} & \multicolumn{2}{c}{Int. time (s)}\\
\colhead{Name} & \multicolumn{2}{c}{(J2000)} & \colhead{J} & \colhead{H} &
\colhead{K$_{s}$} & \colhead{A} & \colhead{ID} & \colhead{Date} & \colhead{SL} &
\colhead{LL}} \startdata
IRAS 04496-6958 & 04 49 18.50  & -69 53 14.3 & 12.66 & 10.85 & 9.43 & 5.61 & 9069312 & 2004-03-01 & 112 & 240 \\
LI-LMC 603 & 05 12 00.82  & -70 32 24.0 & 18.04 & 16.77 & 14.52 & 5.51 & 6078464 & 2004-03-05 & 56 & 56 \\
LI-LMC 1817 & 06 02 45.08 & -67 22 43.0 & 18.43 & 16.23 & 13.53 & 6.08 & 6078208 & 2004-04-14 & 56 & 112 \\
MSX LMC 1205 & 04 51 15.48  & -68 41 40.3 & 14.97 & 12.84 & 11.09 & 7.11 & 11239168 & 2005-03-16 & 48 & 112 \\
MSX LMC 1209 & 04 55 18.88  & -68 46 30.7 & 16.64 & 14.94 & 12.74 & 7.03 & 11239168 & 2005-03-16 & 48 & 112 \\
MSX LMC 1213 & 04 52 20.18  & -68 43 30.1 & 17.98 & 14.85 & 12.44 & 7.08 & 11239168 & 2005-03-16 & 48 & 112 \\
MSX LMC 1298 & 04 51 10.96  & -69 20 14.7 & 15.50 & 13.05 & 11.12 & 5.92 & 11239168 & 2005-03-16 & 48 & 112 \\
MSX LMC 1308 & 04 51 10.96  & -69 20 14.7 & 14.23 & 12.32 & 10.82 & 5.90 & 11239168 & 2005-03-16 & 48 & 112 \\
MSX LMC 1383 & 04 55 33.54  & -69 24 59.3 & 18.11 & 15.85 & 13.08 & 5.90 & 11239424 & 2005-03-14 & 48 & 112 \\
MSX LMC 634 & 05 26 35.70  & -69 08 22.7 & 14.82 & 12.77 & 11.13 & 7.53 & 11238912 & 2005-03-16 & 48 & 112 \\
MSX LMC 768 & 05 31 40.90  & -69 39 19.8 & 15.83 & 13.20 & 11.06 & 5.99 & 11238912 & 2005-03-16 & 48 & 112 \\
MSX LMC 774 & 05 26 23.10  & -69 11 20.3 & 16.78 & 15.08 & 12.72 & 6.26 & 11238912 & 2005-03-16 & 48 & 112 \\
MSX LMC 783 & 05 32 55.47  & -69 20 26.6 & 15.85 & 13.65 & 11.73 & 7.00 & 11238912 & 2005-03-16 & 48 & 112 \\
MSX LMC 787 & 05 34 56.21  & -69 09 16.5 & 16.72 & 14.85 & 12.53 & 7.00 & 11238912 & 2005-03-16 & 48 & 112 \\
MSX LMC 841 & 05 31 13.12  & -66 09 41.2 & 15.78 & 13.40 & 11.55 & 7.07 & 11239680 & 2005-03-16 & 48 & 112 \\
MSX LMC 937 & 05 40 36.06  & -69 52 49.8 & 14.52 & 12.24 & 10.36 & 5.67 & 11239424 & 2005-03-14 & 48 & 112 \\
MSX LMC 950 & 05 35 03.55  & -69 52 45.5 & 14.78 & 12.47 & 10.65 & 6.75 & 11239424 & 2005-03-14 & 48 & 112 \\
MSX LMC 971 & 05 39 51.85  & -70 01 17.0 & 17.84 & 16.48 & 14.21 & 6.75 & 11239424 & 2005-03-14 & 48 & 112 \\
MSX LMC 974 & 05 40 58.18  & -69 53 12.7 & 14.45 & 12.54 & 10.98 & 7.12 & 11239424 & 2005-03-14 & 48 & 112 \\
\enddata
\tablenotetext{a}{\cite{skr06}} \tablenotetext{b}{\cite{ega01}} \tablecomments{Observed
LMC targets sorted alphabetically by the adopted name. RA and Dec coordinates are given
in J2000. Photometry at J, H, and K$_s$ are from 2MASS and the A band (8.3~\um) is from
the MSX survey. We also list the Astronomical Observation Request (AOR) ID along with
date of the observation and integration times for the Short-Low (SL) and Long-Low (LL)
orders in terms of seconds.}
\end{deluxetable*}

The Magellanic Clouds provide excellent laboratories to study the effects of metallicity
on the formation of dust. Since the distances to these galaxies and their metallicities
are well-known, we can derive absolute dust mass-loss rates from observed spectral energy
distributions \citep{leb97,gro07}. On average, the Large Magellanic Cloud's (LMC)
metallicity is about half that of the Sun while the metallicity of the Small Magellanic
Cloud (SMC) is approximately $0.2$ compared to solar \citep{rus89,ber94}. While most
Galactic AGB stars appear to show an oxygen-rich chemistry, the Magellanic Clouds contain
a larger fraction of carbon stars, in line with their lower metallicities \citep{bla80}.

Here we present data from the Infrared Spectrograph \citep[IRS;][]{hou04} aboard the
\emph{Spitzer Space Telescope} \citep{wer04} on 19 carbon-rich AGB stars that we combine
with previously published carbon star samples \citetext{\citealt{slo06,zij06,lag07};
hereafter SZL}. 
The combination of these large samples of Galactic, LMC, and SMC spectra provides us with
a tool to analyze the effects of metallicity on the composition of the dust and gas and
allows us to better understand the dust condensation sequence, the efficiency of grain
formation, and ultimately the mass-loss rates.

\section{Observations}

\subsection{Sample selection and selection effects}

We present \emph{Spitzer} IRS data of 19 carbon-rich AGB stars in the LMC
(Figure~\ref{fig:allplots}; Table~\ref{tbl:obover}). Based on the color classification
from \cite{ega01}, these objects were originally observed for a survey of the dust
condensation sequence in oxygen-rich AGB stars, but their IRS spectral signature clearly
identify them as being carbon rich. This sample covers the same range of ${\rm H}-{\rm
K_s}$ vs ${\rm K_s}-{\rm A}$ color-color space as those LMC objects published by
\cite{zij06} thereby nearly doubling the available data points. The IRS spectrum of IRAS
04496-6958 was previously published by \cite{spe06}.

In order to analyze the effects of metallicity, we compare the combined LMC sample to
C-rich AGB stars observed in the SMC and Milky Way. The SMC stars included here were
observed using the IRS and presented in two separate papers by \cite{slo06} and
\cite{lag07}. We used Galactic data observed with the Infrared Space Observatory (ISO)
and published in the Short-Wavelength Spectrometer (SWS) Atlas \citep{slo03}. Thus, we
present carbon stars from three different metallicity environments: 29 objects residing
in the SMC, 44 from the LMC, and 34 Galactic carbon stars.

The SWS instrument on ISO observed stars previously known to be
carbon stars in the Milky Way with a wide range of colors. Since the
threshold mass to form a carbon star is high in single star
evolution in the Galaxy, not all of these are `normal' AGB stars. In
particular, the blue stars are more peculiar high-mass stars or
binaries \citep{ols75}. Whereas more than half of the Galactic
sample contains blue objects expected to be naked carbon stars, the
LMC and SMC samples are much redder on average. This is a known
selection effect since the Magellanic Cloud surveys were selected
from the mid-IR color classification scheme using 2MASS and MSX
colors \citep{ega01}. Since MSX is only sensitive towards objects
with a strong 8~\um\ excess, it was expected that most of the
observed objects would have high mass-loss rates and relatively red
spectrum. Indeed, for most of the Magellanic Cloud objects, we find
gas mass-loss rates of $10^{-6}$~\smpy\ or higher while Galactic AGB
stars have a wider range of $10^{-8.5}$ to $10^{-4.5}$~\smpy\
(Section~\ref{mloss}). The Galactic sample contains the widest range
of colors spanning between $0.3\lesssim {\rm H}-{\rm K_s}
\lesssim4.5$ while the LMC and SMC objects range between $1\lesssim
{\rm H}-{\rm K_s} \lesssim3.2$ and $0.9\lesssim {\rm H}-{\rm K_s}
\lesssim2.1$ respectively. On average, the carbon stars observed in
the LMC are redder than those observed in the SMC.

\subsection{Spectral features}

\subsubsection{\acet\ absorption}

For the majority of stars in the three samples, the most prominent
molecular absorption features occur at $\sim$7.5 and $\sim$14~\um\
and are normally associated with acetylene (\acet) absorption.
However, HCN and CS absorption are also known to provide
contributions to the depth of these features \citep{goe81,aok98}.
Using the 3.5~\um\ absorption feature, \cite{van06a} show that HCN
and CS are comparatively weak for LMC stars while \cite{mat06} found
no clear evidence of HCN among their LMC sample. In particular, they
interpret the low HCN abundance in the Magellanic Clouds as being
due to initially low nitrogen abundances and efficient carbon
production in metal-poor environments. \cite{aok99} and \cite{cer99}
only detect HCN in 1/3 of their Galactic carbon sample. Thus, we can
confidently identify these two absorptions as prominently due to
\acet.

The double peaked absorption band centered at 7.5~\um\ arises from the \acet\
$\nu^1_4+\nu^1_5$ transitions in the P and R branches. The Q branch band of the \acet\
fundamental $\nu_5$ bending mode creates a sharp absorption feature at 13.7~\um. Colder
gas in the extended envelope (possibly above the dust-forming region) produces the sharp
Q-branch feature while the broad P and R branches originate from warmer gas deeper into
the atmosphere \citep{jor00}. \cite{mat06} explain the broad wings extending on either
side of the 13.7~\um\ feature as being due to the P and R branches. In the most extreme
cases, these wings can extend between $\sim$11 and $\sim$17~\um\ \citep{spe06}.
Throughout this paper, we refer to the Q branch as the ``13.7~\um\ feature'' and the P+R
branches in the same region as the ``14~\um\ feature''.

\subsubsection{10-micron absorption}

A significant number of stars exhibit a relatively weak absorption
band spanning $\sim$8.5 to $\sim$11~\um\ with a peak depth around
9.4~\um. We mainly observe this feature in the bluest objects.
While the carrier of this feature is still in doubt, our
observations support a molecule that originates near the photosphere
of the star and would be masked by the infilling of dust emission in
redder objects.

\cite{jor00} and \cite{zij06} favor an identification of C$_3$;
however, we do not observe a corresponding absorption between 6.6
and 7~\um\ that is also predicted. Furthermore, we cannot use the
deep absorption feature at 5~\um\ to confirm the presence of this
molecule since this band also contains significant contributions
from CO. Laboratory measurements are necessary in order to firmly
identify C$_3$ as the carrier of the 10-micron absorption.

\subsubsection{SiC emission}

The majority of the carbon stars in the Magellanic Clouds and the Galaxy show the
presence of an emission feature around 11.3~\um\ generally attributed to SiC emission
\citep{tre74}. While the central wavelength of the SiC resonance as measured in the
laboratory intrinsically lies at 11.3~\um\ \citep{gil71}, the central peak in
astronomical data varies between 11.15 and 11.7~\um\ for Galactic sources. \cite{spe05}
used grains of different sizes to interpret this shift: larger grains provide a cool
absorbing component around 10.8~\um\ while a warmer emitting component towards the red
arises from smaller grains. Thus, different grain-size distributions will theoretically
provide different contributions from emitting and absorbing components, causing the
overall central wavelength of the SiC feature to shift.

\subsubsection{MgS emission}

In slightly less than half of all the objects presented here, we
observe a broad emission feature beginning blueward of 22~\um\ and
extending beyond the IRS spectral range. This feature is commonly
seen in carbon-rich AGB and post-AGB stars as well as planetary
nebulae and consists of two subpeaks at 26 and 33 \um\ \citep{for81}
both of which are attributed to magnesium sulfide \citep[MgS;
][]{goe85}. By calculating absorption cross sections for various
grain shapes, \cite{hon02} modeled the MgS spectral feature of
Galactic carbon stars and showed that the subpeaks are due to
different distributions of grains shapes and that the peak
wavelength of the feature shifts with temperature.

\section{Analysis}

\subsection{Manchester method}

\begin{figure}[b]
\begin{center}
\includegraphics[width=.3\textwidth, angle=90]{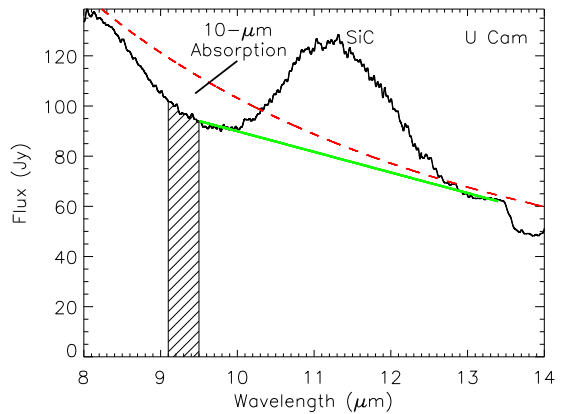}\\
\caption{{The spectra of U Cam showing the effect of the 10~\um\ absorption feature on
the [9.3] Manchester band represented as the cross-hatched vertical bar. The 10~\um\ band
gives a flux value significantly below the probable continuum level, which is
approximated by a modified black body. The resulting straight line continuum fit (solid
line) differs significantly from the modified blackbody fit (dashed line) underneath the
SiC emission.}} \label{fig:c3man}
\end{center}
\end{figure}

Since the dominant component of the dust in these sources consists of featureless
amorphous carbon \citep{mar87}, it is useful to develop an observable quantity that
represents the dust opacity. Previous analyses of carbon star data have utilized the
``Manchester method'' \citep{slo06} in order to define the amorphous carbon dust
continuum amid the plentiful spectral emission and absorption features. This method
defines four narrow bands centered at 6.4, 9.3, 16.5, and 21.5~\um\ that are used to
determine a set of two color temperatures of \ctemp\ and $[16.5]-[21.5]$. These colors
can then be used as proxies for the optical depth of the surrounding dust shell
($\tau\propto[6.4]-[9.3]$) and for estimating the dust temperature, respectively.

\begin{figure}[htp]
\begin{center}
\includegraphics[width=.6\textwidth, angle=90]{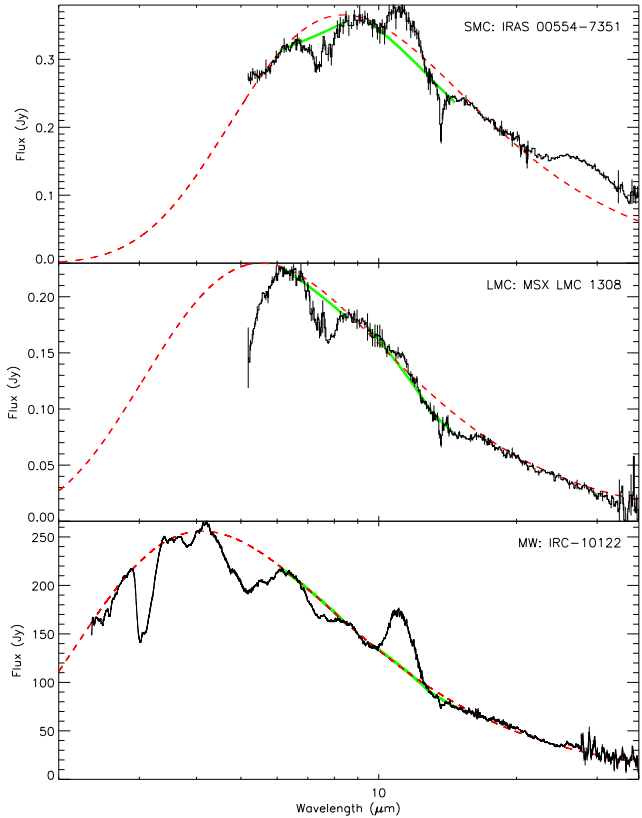}\\
\caption{{Example spectra with error bars of three sources, one from each environment --
SMC, LMC, and the Milky Way from top to bottom, respectively. The dotted line indicates
the chi-square fits of the underlying carbon-rich dust continuum using a
single-temperature modified blackbody. The straight line segments represent the estimated
continuums using the method employed by SZL.
We find the feature strengths by dividing the derived blackbody fits.}} \label{fig:3spec}
\end{center}
\end{figure}

A few complications arise when trying to perform this method on a sample of objects
containing a wide range of dust temperatures. In particular, the Manchester 9.3~\um\ band
fall directly within the 10~\um\ absorption feature, thus underestimating the dust
continuum flux at this wavelength for stars with this feature (Figure~\ref{fig:c3man}).
However, this only affects a small number of the bluest objects providing a mean shift of
0.07 for the \ctemp\ color. Thus, we still rely on the \ctemp\ color to provide us with a
qualitative analysis of the dust opacities. Furthermore, the 21.5~\um\ band, used to
define the blackbody continuum, can acquire increased flux from the adjoining broad MgS
feature for significant emission strengths, causing the continuum emission to be
overestimated at these wavelengths. Prominent MgS emission can occur shortward of 20~\um\
for low dust temperatures \citep{hon02}.

Additionally the Manchester method adopts straight-line continuum
fits in order to measure the strengths of the \acet\ absorption
bands and the SiC emission feature. They defined wavelength segments
of supposed continuum that flank the features of interest
\citep[see][]{slo06}. 
However, since the blue side of the continuum used for the SiC
feature ranges from 9.5 to 10.1~\um, this method will tend to
underestimate the underlying continuum and overestimate the flux in
the SiC band (Figure~\ref{fig:c3man}). Similarly, the varying
strength of the \acet\ feature affects determination of the red side
of the SiC feature.
In the most extreme cases, \acet\ absorption can reach as low as
$\sim$11~\um\ causing considerable suppression of the observed SiC
emission. The same variation in broadness of the \acet\ bands also
makes it difficult to set a consistent wavelength range at 7.5 and
14~\um\ that allows for proper isolation of the P, Q, and R branches
of \acet. For this reason, previous studies only focused on the
narrow Q branch of the 13.7~\um\ feature and disregarded the
flanking P and R branches. While this paper strives to improve on
quantitative determinations of the various spectral features, we
stress that our improvements described in the following section do
not change the majority of the qualitative conclusions found by SZL
using the Manchester method.

\subsection{Modified blackbody fits}

\begin{figure}[b]
\begin{center}
\includegraphics[width=.3\textwidth, angle=90]{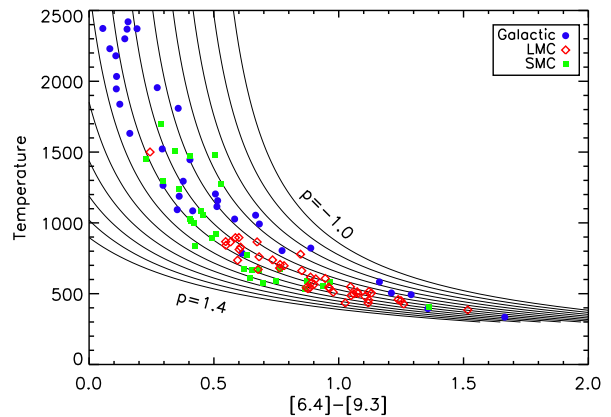}\\
\caption{{Contours of constant p-parameter values for modified blackbodies with
temperatures ranging between 300 and 2500~K with $-1 \leq p \leq 1.4$ with a 0.2
increment. The symbols represent data from fitted blackbodies for the samples of
Galactic, LMC, and SMC objects. Error bars for \ctemp\ are on the order of the symbol
size.}} \label{fig:ctcorr}
\end{center}
\end{figure}

SZL employed the Manchester $[16.5]-[21.5]$ color to derive a blackbody color temperature
and thereby extrapolate the continuum underneath the MgS feature. However, the
aforementioned problems associated with this type of continuum estimation can lead to
inaccurate determinations of dust properties such as the temperature, mass-loss rate, and
total dust mass. We improve on this method by fitting a modified blackbody to each carbon
star's spectrum. Expanding on the method described by \cite{hon02}, we utilize a $\chi^2$
fitting routine to find the best single temperature modified blackbody to model the
entire spectral continuum,
\begin{equation}\label{modbb}
    F(\lambda)=A \times B(\lambda,T) \times \lambda^{-p},
\end{equation}
where $F(\lambda)$ is the flux density of the continuum at some wavelength $\lambda$,
$B(\lambda,T)$ is the Planck function with a temperature $T$, $p$ is the dust emissivity
index, and $A$ is a scaling factor. We minimize the $\chi^2$ values by varying the free
parameters ($T$, $p$, and $A$) from Equation~\ref{modbb} over selected continuum points.
The $p$-parameter represents how efficiently the dust grains radiate at wavelengths
larger than the grain size. To some extent, varying this parameter allows us to
incorporate effects from optical depth and temperature gradients. Crystalline material
has $p\sim2$, amorphous material has an emissivity index between 1 and 2, and layered
materials have $p\sim1$. Temperature gradients within the dust shell lower the value of
$p$ by broadening the spectral energy distribution. Similarly, an optically thick dust
shell further reduce the value of $p$.

\begin{figure}[tp]
\begin{center}
\includegraphics[width=.3\textwidth, angle=90]{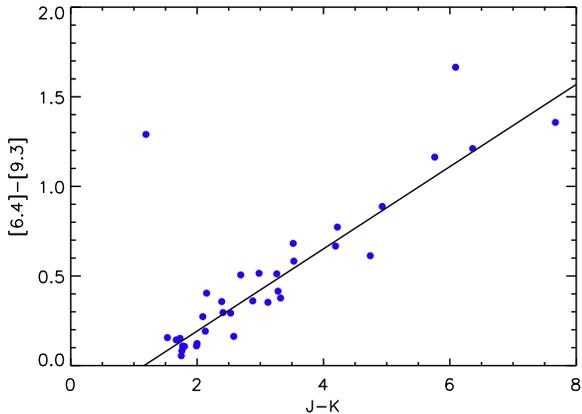}\\
\caption{{Correlation between 2MASS ${\rm J}-{\rm K}$ and \ctemp\ colors for Galactic
objects. We have overplotted the best-fit linear relation.}} \label{fig:tmcorr}
\end{center}
\end{figure}

For a source embedded in a circumstellar shell, one expects the continuum emission to
follow the general form of
\begin{equation}\label{agbflux}
    F_{\nu}(\lambda)=I_{\nu}(\lambda,T_{\star}) e^{-\tau_{\nu}} + B_{\nu}(\lambda,T_{CS}) (1-e^{-\tau_{\nu}}),
\end{equation}
where $I_{\nu}(\lambda,T_{\star})$ is the stellar emission, $d$ is the distance to the
source, $B_{\nu}(\lambda,T_{CS})$ is the blackbody emission of the circumstellar dust
shell, $\tau_{\nu}$ is the optical depth of the dust shell, and $T_{\star}$ and $T_{CS}$
are the prevailing temperatures of the stellar photosphere and envelope, respectively. In
cases where the optical depth is large at infrared wavelengths, the contribution from the
stellar photosphere in Equation~\ref{agbflux} is negligible, and the modified blackbody
fit (Equation~\ref{modbb}) gives an approximation for the average temperature of the dust
shell. However when the optical depth is small, the photospheric emission becomes
important, and both $T_{\star}$ and $T_{CS}$ contribute to the temperature in
Equation~\ref{modbb}. For this case, the temperature $T$ from the modified blackbody fit
should be treated only as a fit parameter.

The blackbody continuum fits allow us to improve on straight line continua by determining
more precisely the wavelength range over which spectral features are present
(Figure~\ref{fig:3spec}).
In order to accommodate for the varying widths of the features, the continuum points were
selected for each spectrum individually. We chose continuum points redward of the
7.5~\um\ \acet\ band at; between the 7.5~\um\ \acet\ band and SiC emission feature; and
between the 14~\um\ \acet\ and MgS feature, taking care that none of the feature wings
were included in the continuum selection. When fitting ISO spectra, the inclusion of
wavelengths shortward of 5.2~\um\ that are unavailable in IRS spectra alters the best-fit
temperature up to 10\% and slightly shifts $p$ towards negative values. In the majority
of these cases, though, only minor differences arise in the temperature and p-value when
considering the 2.3 to 5.2~\um\ region to constrain the continuum. Moreover, changes to
the fitted continuum above and below the features of interest were imperceptible;
therefore, we do not introduce a bias in the feature strengths of Galactic objects.

In addition, we considered 2MASS J, H, and K$_{\rm s}$ points \citep{skr06} to constrain
the shorter wavelengths of the blackbody points. However, due to the intrinsic
variability of these objects, the 2MASS data points were only used to guide the slope of
continuum fit rather than as rigid anchors. Additionally, different dust temperatures
dominate the emission between the near-IR 2MASS photometry and the mid-IR spectra, but we
partially account for this effect with the inclusion of the dust emissivity index, $p$.
Also, due to the relative distance and faintness of objects in the Magellanic Clouds as
compared to objects in the Galaxy, spectral and photometric data for LMC and SMC objects
contain larger uncertainties than Galactic objects. In the entire sample, 28 objects have
a J-band magnitude greater than the 10-$\sigma$ point-source detection limit of 15.8~mag,
with 23 of these objects residing in the LMC. This number improves by
half for the 10-$\sigma$ H-band limit of 15.1~mag. 
Thus, using the 2MASS points as absolute constraints for the
single-temperature blackbody fits was not recommended in every case.

\begin{figure}[b]
\begin{center}
\includegraphics[height=0.45\textwidth, angle=90]{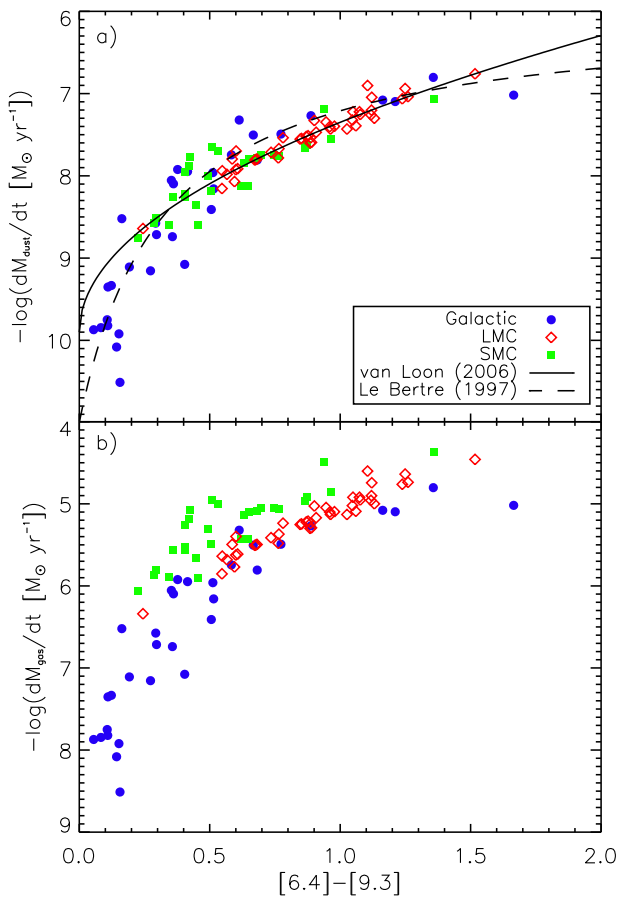}\\
\caption{{Plots of the mass-loss rates derived from Equations~\ref{mdotlmc} through
\ref{mdotmw} with respect to the \ctemp\ color. The top plot shows that $\dot{M}_{dust}$
derived for the objects in the Milky Way from ${\rm J}-{\rm K}$ \citep{leb97} correlates
with those approximated from the \ctemp\ color \citep{van07}. Due to difference in
assumed dust-to-gas ratios, each metallicity environment has a slightly different
correlation between the gas mass-loss rate and the \ctemp\ color.}} \label{fig:mloss}
\end{center}
\end{figure}

\begin{figure*}[t]
\begin{center}
\includegraphics[height=0.9\textwidth, angle=90]{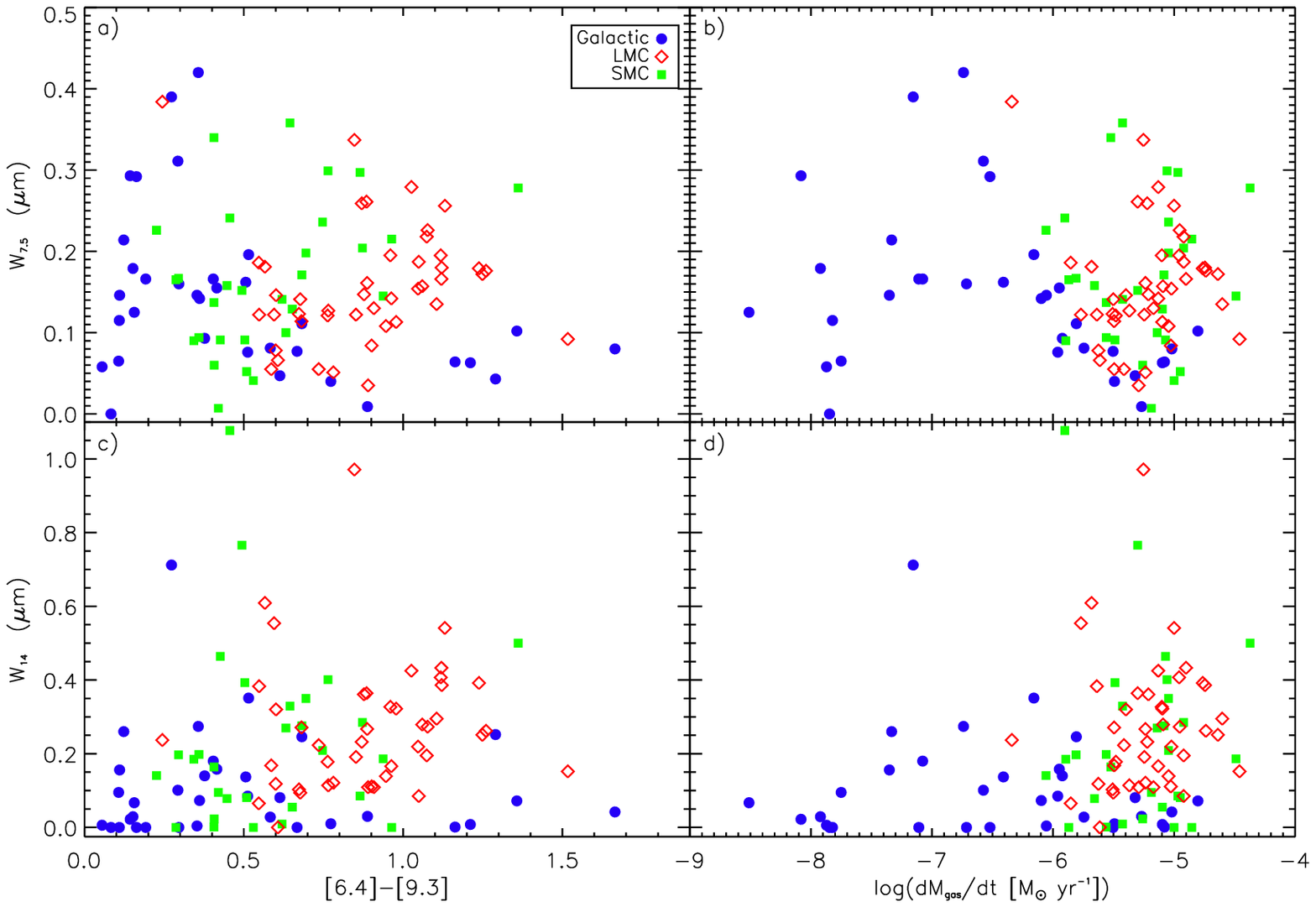}\\
\caption{{From left to right, the top panels plot the equivalent widths for the \acet\ P
and R branches at 7.5~\um\ as a function of \ctemp\ color (dust mass-loss rate) and gas
mass-loss rates, respectively. The bottom panels show the equivalent widths for \acet\ P
and R branches at 14~\um\ as a function of \ctemp\ and gas mass-loss rates.}}
\label{fig:ew}
\end{center}
\end{figure*}

Figure~\ref{fig:ctcorr} shows the result of Equation~\ref{modbb} for modified blackbodies
ranging from temperatures of 300 to 2500~K and p-values of $-1.0$ to $1.4$. Continuous
lines show how the temperature changes for varying \ctemp\ with constant p-values. We
overplot the derived p and T values against the measured \ctemp\ values for the SMC, LMC,
and Galactic samples. The mean p-values for the entire sample is $0.13\pm0.26$, and for
the individual samples are $0.27\pm0.2$ (SMC), $0.18\pm0.22$ (LMC), and $-0.05\pm0.25$
(Galactic). The variation in average values between the three environments may indicate a
metallicity effect; however, since the p-parameter represents the optical properties of
the grain which evolve over the lifetime of these objects, this may reflect the biases of
each sample. For instance, the objects within the Milky Way contain the widest range of
colors with the majority residing towards the blue, while the SMC sample contains the
smallest color range and consists of much redder objects.

Due to the poor quality of some spectra, mis-pointings, and/or
background contamination, we were unable to obtain blackbody fits
for a total 11 objects from the various samples: MSX~LMC~782 from
our sample; NGC~1978~MIR1, NGC~1978~IR4 \citep{zij06};
NGC~419~LE~18, NGC~419~LE~27, NGC~419~LE~35, and ISO~00549
\citep{lag07}; and IRAS~18430-0237, S~Lyr, SS73~38, and SZ~Sgr
\citep{slo03}.

Table 2 lists the feature strengths determined using our modified blackbody fits. We
measure equivalent widths of the 7.3~\um\ band, 13.7~\um\ \acet\ Q branch, and the P+R
branches at 14~\um. Strengths of the dust emission features were determined by taking the
ratio of the integrated flux to the integrated continuum. The central wavelength for the
SiC feature was also found. We do not include features where low signal-to-noise ratios
cause uncertainties comparable to the strength.

\subsection{Mass-loss Rates \label{mloss}}

\begin{figure*}[htp]
\begin{center}
\includegraphics[height=0.9\textwidth, angle=90]{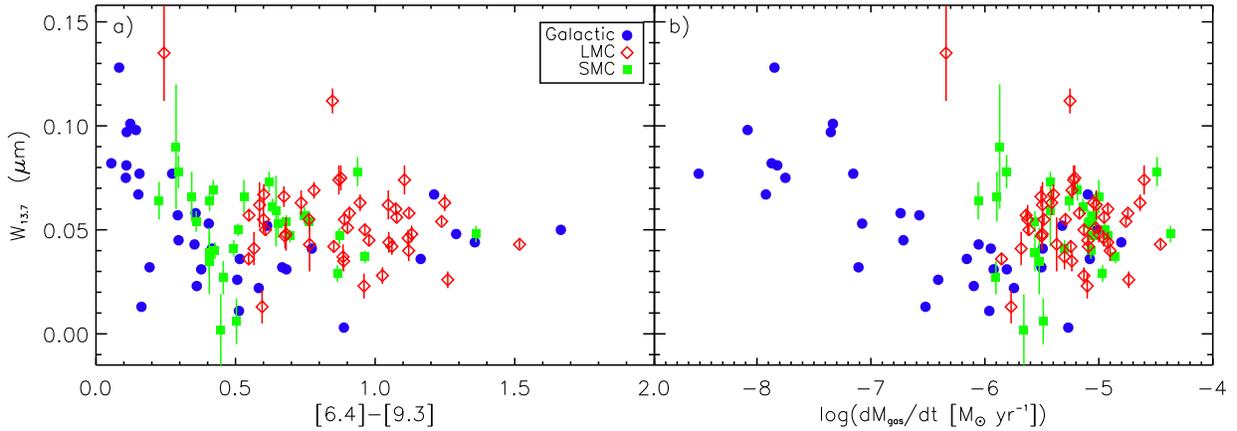}\\
\caption{{\textbf{(a)}: Plot of the equivalent with for the 13.7~\um\ \acet\ Q branch as
a function of the \ctemp\ color (dust mass-loss rate). \textbf{(b)}: The \acet\
equivalent width for the 13.7~\um\ Q branch as a function of the derived gas mass-loss
rate.}} \label{fig:ew137}
\end{center}
\end{figure*}

Based on dust radiative transfer models, \cite{gro07} determined dust and gas mass-loss
rates for 60 carbon stars in the Magellanic Clouds while \cite{leb97} modeled the
circumstellar dust shells of 23 Galactic carbon stars. These works derived
$\dot{M}_{dust}$ directly from the infrared SEDs and NIR photometry. Gas mass-loss rates
were subsequently derived by scaling by an assumed dust-to-gas ratio ($\psi$) based on
the metallicity. Even though this is a common assumption when deriving $\dot{M}_{gas}$,
there are no studies that conclusively show that the total mass-loss rate explicitly
depends on metallicity \citep{zij96}. The dust-to-gas ratio has been shown to depend
approximately linearly on initial metallicity while mass-loss rate shows only a weak
dependence \citep{van00}. However, in order to proceed with our analysis of the
dependencies of dust and molecular abundances with pressures and densities, we must make
some assumption about the gas mass-loss rates.

When available, we use mass-loss rates published by \cite{gro07} or \cite{leb97}. In
their derivation of $\dot{M}_{gas}$, \cite{gro07} used the same dust-to-gas ratio of
$\psi=0.005$ for all objects within the LMC and SMC. Instead, we have adopted
$\psi=0.002$ for SMC objects, corresponding to a metallicity of $Z/Z_{\sun}=\frac{1}{5}$.
This assumes that $\psi$ scales with metallicity and $\psi_{MW}=0.01$.

For the objects within Magellanic Clouds, \cite{gro07} show a correlation between
$\dot{M}_{dust}$ and the \ctemp\ color (Figure~\ref{fig:mloss}a). The gas mass-loss rate
for the LMC objects can then be derived by \citep{van07}:
\begin{equation}\label{mdotlmc}
  \log(\dot{M}_{gas,LMC})=2.55([6.4]-[9.3])^{0.5}-7.6.
\end{equation}
For the SMC sample, we modify Equation~\ref{mdotlmc} using a value of $\psi_{SMC}=0.002$,
\begin{equation}\label{mdotsmc}
  \log(\dot{M}_{gas,SMC})=2.55([6.4]-[9.3])^{0.5}-7.2.
\end{equation}
All mass-loss rates are in ${\rm M_{\sun}~yr^{-1}}$.

Since these relations were derived only for objects between
$0.2\leq[6.4]-[9.3]\leq1.5$, the mass-loss rates of objects with
$[6.4]-[9.3]<0.2$ are not well constrained, affecting 11 Galactic
objects. Instead, we determined gas mass-loss for all Galactic
objects using the relation from \cite{leb97},
\begin{equation}\label{mdotmw}
     \log(\dot{M}_{gas, MW})=-6.0/({\rm J}-{\rm K}-0.2)-4.0,
\end{equation}
assuming a dust-to-gas ratio of 0.01. We find a linear correlation between the ${\rm
J}-{\rm K}$ and \ctemp\ colors for Galactic objects (Figure~\ref{fig:tmcorr}). Therefore,
we can overplot the trend from \cite{leb97} in terms of the \ctemp\ color. The top panel
of Figure~\ref{fig:mloss} shows that the dust mass-loss rates of the Galactic objects
derived from 2MASS ${\rm J}-{\rm K}$ indeed overlap with those derived from the \ctemp\
color. Once we apply metallicity corrections for determining the gas mass-loss rates,
objects in the separate samples no longer fall along a single trend
(Figure~\ref{fig:mloss}b). Table~\ref{tbl:mloss} lists the dust and gas mass-loss rates
for each object.

\section{Results}

\subsection{Acetylene molecular absorption}

\subsubsection{P \& R branches}

\begin{figure*}
\begin{center}
\includegraphics[height=0.9\textwidth, angle=90]{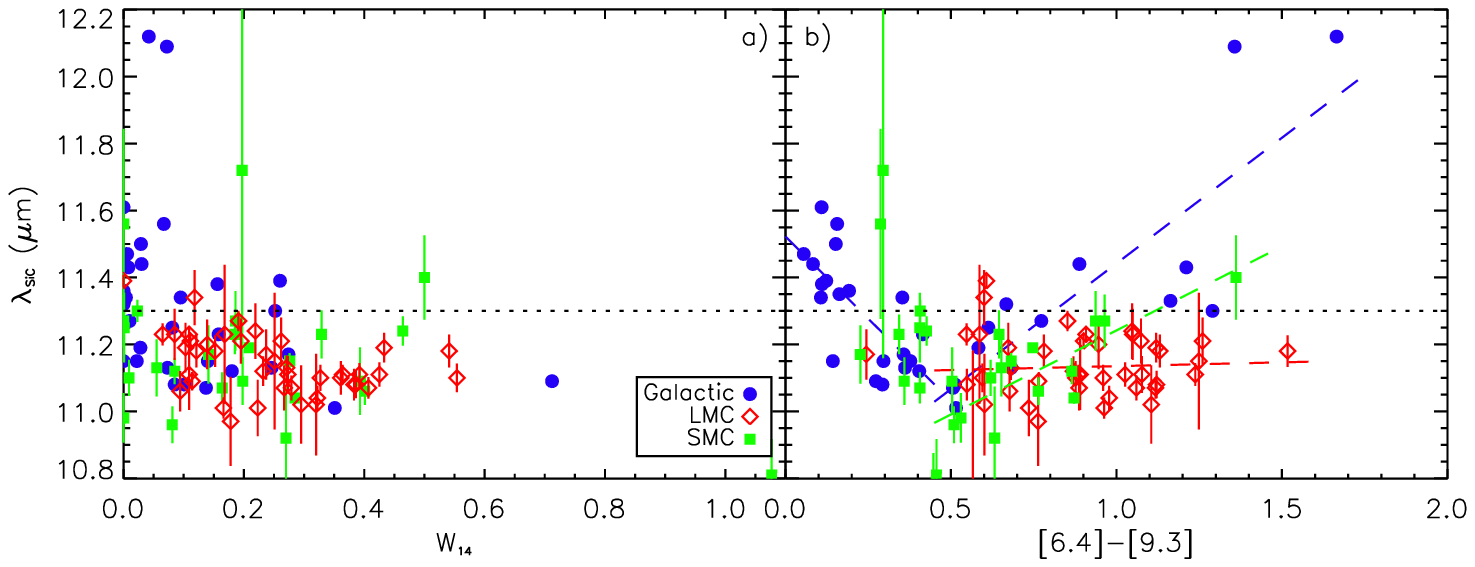}\\
\caption{{\textbf{(a)}: Correlation between the 14 \um\ equivalent width and SiC central
wavelength. Error bars for Galactic objects are smaller than symbols. Objects with
increasingly large \acet\ band absorptions tend to have $\lambda_{SiC}$ shifted further
towards the blue while objects with redder $\lambda_{SiC}$ have smaller \acet\ equivalent
widths. \textbf{(b)}: The SiC apparent central wavelength as a function of the \ctemp\
color similar to Figures 9 \& 10 from \cite{slo06}. The dashed lines are linear fits to
trends described in the text. The horizontal dotted line indicates the SiC central
wavelength observed in laboratory data.}} \label{fig:censic}
\end{center}
\end{figure*}

We plot the equivalent widths of the \acet\ bands at 7.5 and 14~\um\ as a function of the
\ctemp\ color and $\dot{M}_{gas}$ for the Galactic, LMC, and SMC samples in
Figure~\ref{fig:ew}. By plotting against the \ctemp\ color, we are comparing sources with
the same optical depth. Assuming a constant expansion velocity independent of
metallicity, the \ctemp\ color also directly correlates with the dust mass-loss rate
assuming \citep{gro07}. \cite{mat02} show that increasing the contribution from dust
emission can affect the molecular band strength by filling in the \acet\ absorption and
decreasing the equivalent width. \cite{mat05} and \cite{van06a} show evidence of this
``dust dilution'' for \acet\ bands occurring at 3.1 and 3.8~\um\ in the Magellanic
Clouds.

Analyzing the Galactic sample, we find that the 7.5~\um\ band shows a behavior similar to
the 3.8~\um\ \acet\ bands from \cite{mat05}. Figures~\ref{fig:ew}a and \ref{fig:ew}b show
an initial increase in equivalent width with optical depth and mass-loss rate, indicating
a rise in level populations from increasing pressure and density. Circumstellar emission
subsequently fills in the absorption band at $\dot{M}_{gas}\sim10^{-6.5}$ causing a
decline in the equivalent width. The 7.5~\um\ absorption feature reflects this trend
better than the 14~\um\ band since the SiC emission feature tends to overlap the 14~\um\
P branch.

The P and R branch acetylene bands also show a dependence on metallicity. For a given
\ctemp\ color, objects from the SMC sample show slightly larger equivalent widths than
those from the LMC. This is even more pronounced when compared to Galactic sources at 7.5
and 14~\um\, albeit with considerable scatter (Figures~\ref{fig:ew}a and \ref{fig:ew}c).
However, plotted against gas mass-loss rate, those objects from the LMC and SMC occupy
the same space (Figures~\ref{fig:ew}b and \ref{fig:ew}d). Our assumptions when deriving
gas mass-loss rates effectively discard the metallicity dependence between the LMC and
SMC objects found by \cite{lag07} when using different and equally valid assumptions.
Thus, we must be aware that the assumptions we make affect qualitative results when
dealing with the gas mass-loss rates. However, the LMC and SMC objects show significantly
higher equivalent widths than their Galactic counterparts with the same $\dot{M}_{gas}$
as found by SZL. These works suggest that efficient dredge-ups for low metallicity
C-stars help to increase the C/O ratio and, therefore, the \acet\ equivalent width for
these bands. Additionally, for a given $\dot{M}_{gas}$, objects in the Magellanic Clouds
will contain less dust than Galactic objects causing effects from dust dilution to be
less pronounced for the SMC and LMC samples.

\subsubsection{13.7 \um\ Q branch}

Figure~\ref{fig:ew137} shows the equivalent width of the Q branch band at 13.7~\um\ as a
function of the \ctemp\ color and $\dot{M}_{gas}$. For Galactic objects, we observe a
trend of the 13.7~\um\ Q branch equivalent width with mass-loss rate. The equivalent
width initially decreases until $[6.4]-[9.3]=0.5$ and $\dot{M}_{gas}=10^{-6}$~\smpy\ then
very slightly increases. Van Loon et al.\ (2006) observe similar trends in absorption
bands at 3.1, 3.57, 3.7, and 3.8 \um\ and explain them as being due to a competition
between the effect of dust dilution and the effect of higher densities from an increasing
mass-loss rate. The initial decline towards red colors in the Galactic sample suggests
that circumstellar dust emission fills in the absorption feature. As gas mass-loss rate
and pressure increase, densities become sufficiently high to overcome effects from the
veiling at around $10^{-6}$~\smpy\ and effectively increase the 13.7~\um\ equivalent
width. This would suggest that the 13.7~\um~\acet\ band has a stronger dependence on the
gas density than the dust column density at higher $\dot{M}_{gas}$.

We observe a similar trend of initially decreasing 13.7~\um\ equivalent width for SMC
objects in Figure~\ref{fig:ew137}a at $[6.4]-[9.3]<0.5$, which overlaps the Galactic
trend. However, when plotted against $\dot{M}_{gas}$, these particular objects no longer
correspond to the observed Galactic trend, supporting the interpretation that dust
dilution has a prominent influence on the 13.7~\um\ absorption at lower mass-loss rates.
Such a trend is not observed for the LMC sample since only one object (GRRV~38) has a
value of \ctemp\ less than 0.5. At $\dot{M}_{gas}\gtrsim10^{-6}$~\smpy, though, the
slightly increasing trend of the Galactic objects is generally matched by both the LMC
and SMC objects. This further suggests that effects from increasing densities dominate
over effects from dust dilution at higher gas mass-loss rates for this absorption band.

Figure~5 of \cite{mat06} show the same increase in the 13.7 \um\ equivalent width with
\ctemp\ color and interpret this as evidence for a lack of dust dilution. However, except
for one star, their entire sample consists of objects redder than $[6.4]-[9.3]=0.5$ with
$\dot{M}_{gas}\gtrsim10^{-5.5}$~\smpy; therefore, they were unable to observe the initial
decrease in the equivalent width. Although we observe the more obvious effects of dust
dilution at lower optical depths and lower $\dot{M}$, we still support the conclusions of
\cite{mat06} that the molecules exist throughout the envelope. Since the equivalent
widths do not continue to decline at higher mass-loss rates where photospheric molecules
would be veiled by the overlying dust, then the absorbing molecules most likely exist
throughout the circumstellar envelope.


\subsection{Dust emission}

\subsubsection{SiC central wavelength}

\begin{figure*}[htp]
\begin{center}
\includegraphics[height=0.9\textwidth, angle=90]{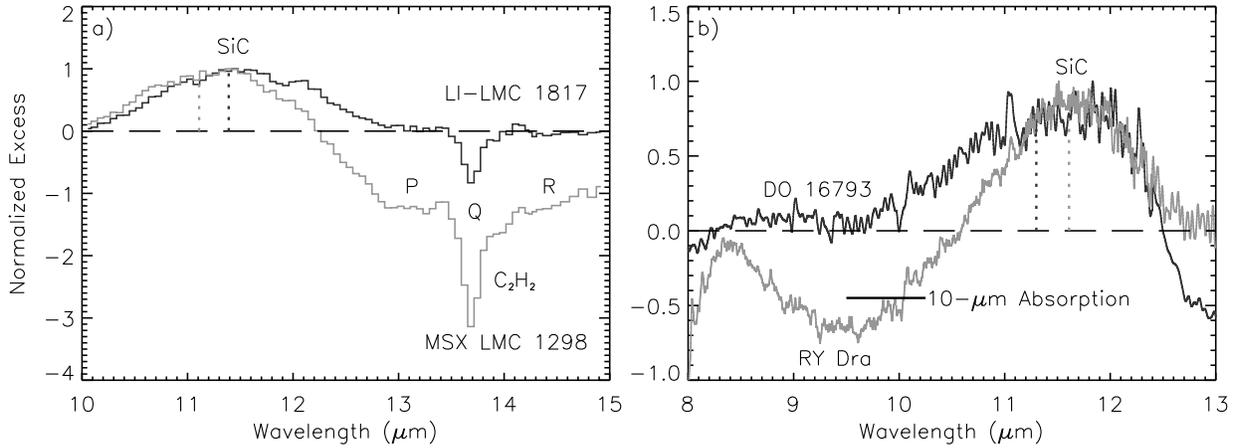}\\
\caption{{\textbf{(a)}: The dark line shows the normalized continuum subtracted spectrum
of LI-LMC~1817. Overplotted in gray is the spectrum of MSX~LMC~1298. \acet\ absorption
including P, Q, and R branches are labeled as well as the SiC emission feature. The
vertical dotted lines indicate the centroid of the SiC feature as it is affected by
\acet\ absorption. \textbf{(b)}: The spectrum of DO~16793 is plotted in black while
RY~Dra is in gray. The 10~\um\ absorption and SiC emission features are labeled, and the
vertical dashed lines show the shift of $\lambda_{SiC}$ when affected by the absorption
band.}} \label{fig:c3sic}
\end{center}
\end{figure*}

The central wavelength of the SiC feature ($\lambda_{SiC}$) has been used to infer dust
properties and circumstellar geometry \citep{spe05}, but is very sensitive to the
placement of the continuum. Despite the improved continuum determination, spectra with
prominent acetylene bands at 14~\um\ continue to affect the central wavelength by
removing emission from the SiC feature (Figure~\ref{fig:censic}a). Objects with larger
$\lambda_{SiC}$ tend to have little \acet\ absorption while objects with significant
\acet\ absorption have $\lambda_{SiC}$ occurring towards the blue. \cite{zij06} and
\cite{lag07} raise the possibility that since the \acet\ absorption overlaps the SiC
emission, the red wing of the SiC feature is suppressed and causes the central wavelength
to shift towards shorter wavelengths (see also Figure~\ref{fig:c3sic}a).

\acet\ suppression can explain the initially decreasing trend for Galactic stars when
plotting \ctemp\ against $\lambda_{SiC}$ reported by \cite{slo06} and displayed in
Figure~\ref{fig:censic}b. Most of the blue Galactic objects have little \acet\
absorption; therefore, the SiC feature has not been altered on the long-wavelength side.
At the same time, absorption at $\sim$10~\um\ lowers emission from the blue side of the
SiC feature, as shown in the Figure~\ref{fig:c3sic}b. This has the opposite effect of the
\acet\ absorption, causing $\lambda_{SiC}$ to be artificially shifted to longer
wavelengths as the temperature increases (\ctemp\ color decreases). Thus, for the bluest
objects, we observe $\lambda_{SiC}$ to be redder than the lab data.

For $[6.4]-[9.3]\gtrsim 0.4$, Galactic carbon stars show increasing $\lambda_{SiC}$ with
increasing values of \ctemp. SMC stars appear to follow the same general trend as the
Galactic sample over the observed color range, albeit a little less pronounced
(Figure~\ref{fig:censic}b). However, the sample of SMC objects contains fewer stars
overall and spans a much smaller color range. LMC stars, on the other hand, show a fairly
consistent central wavelength between 11.1 and 11.2~\um\ for $[6.4]-[9.3]=0.5$ to 1.5.
These objects contain some of the most substantial 14~\um\ \acet\ bands which
dramatically affects the observed SiC central wavelength.

The different trends of the SiC central wavelength observed for the LMC, SMC, and
Galactic samples at $[6.4]-[9.3]\gtrsim0.4$ can all be explained by a competition between
self-absorption of the SiC feature and effects from \acet\ absorption. SiC
self-absorption shifts the feature from 11.3~\um\ towards the red while the broad 14~\um\
\acet\ absorption band causes the SiC feature to shift towards the blue. Hence, objects
containing little \acet\ absorption show a steeper slope with increasing optical depth
(ie. Galactic sample). Since the Magellanic Cloud samples contain significantly wider
\acet\ bands, the slope of this trend decreases.

Two objects, AFGL~341 and IRAS~21489+5201, stand out with extremely
red central wavelengths at about 12.1~\um. These particular objects
are among the reddest sources, exhibit no apparent 10~\um\
absorption feature, and contain a relatively weak \acet\ absorption
at 14~\um. Thus, self-absorption of the SiC feature most likely
produces the considerable shifts in the central wavelength for these
two objects.

While self-absorption of the SiC feature may have a considerable effect on trends of the
Galactic sources, it is difficult to make a conclusive statements due to the molecular
absorptions flanking the dust feature. We find only two objects in our entire sample that
exhibit a shift in $\lambda_{SiC}$ mainly due to SiC self-absorption. Therefore, one must
be careful when making claims about the state of the SiC optical depth and
self-absorption based purely on the position of its central wavelength, especially in the
presence of strong molecular bands.

\subsubsection{SiC feature strength \label{sic}}

\begin{figure*}[htp]
\begin{center}
\includegraphics[height=0.9\textwidth, angle=90]{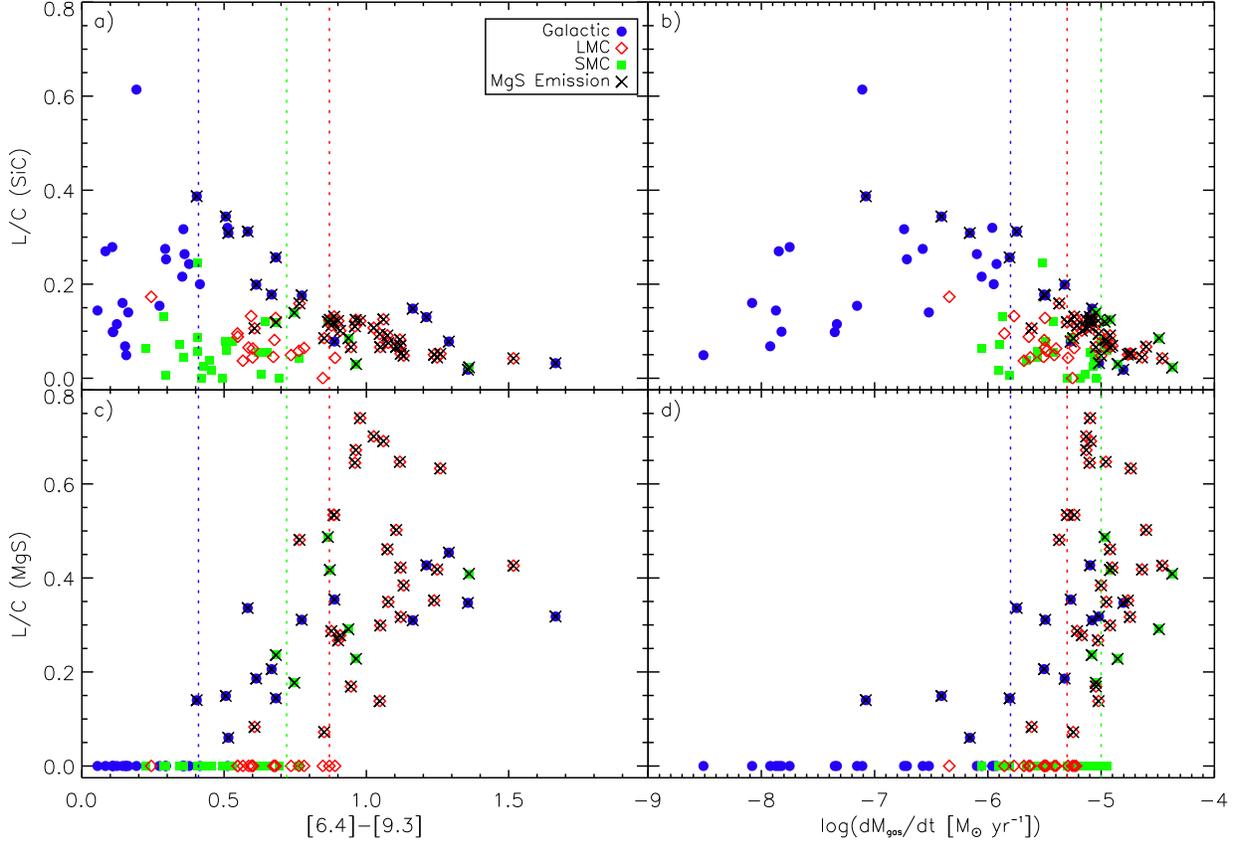}\\
\caption{{Plots of the SiC and MgS dust feature strengths with respect to the \ctemp\
color (dust mass-loss rate) and the gas mass-loss rate. Objects with a measurable MgS
emission feature have been marked in the top two panels with a black X. In general,
redder sources have a weaker SiC emission and correlate with the presence of MgS. Dotted
lines in all panels indicate the transition point in each metallicity system from objects
without any MgS to objects with MgS emission.}} \label{fig:dust}
\end{center}
\end{figure*}

For a given \ctemp\ color (which correlates with dust optical depth and
$\dot{M}_{dust}$), the line-to-continuum ratio of the SiC emission feature appears
largest for Galactic objects, but shows no appreciable difference between the LMC and SMC
objects over their overlapping color range of $0.5<[6.4]-[9.3]<1.0$
(Figure~\ref{fig:dust}a). Plotting the SiC strength against the gas mass-loss rate
provides a good indication of the observational bias inherent in the Magellanic Cloud
samples (Figure~\ref{fig:dust}b). GRRV~38 and RAW~960 are the only two sources within the
Magellanic Clouds with $\dot{M}_{gas}\leq10^{-6}$ \smpy, while about 20 Galactic objects
span the range between $10^{-9}$ and $10^{-6}$ \smpy.

While trends within the Magellanic Clouds are difficult to discern
due to the large scatter of the SMC sources and small number of LMC
sources with $[6.4]-[9.3] \lesssim 0.5$, the Galactic sources
increase rapidly and peak at $[6.4]-[9.3] \sim 0.5$, matching the
observations by \cite{slo06}. This same trend also transfers to
Figure~\ref{fig:dust}b with a general increase in SiC emission
relative to carbon emission from low ($10^{-9}$~\smpy) to
intermediate ($10^{-6.5}$~\smpy) mass-loss.

Moving towards redder colors and higher mass-loss rates, the SiC strength subsequently
decreases. The starting point of this decline appears to depend on metallicity. This
trend is most pronounced in the Galactic objects which initially increases, then
plateaus, and finally decreases between $10^{-9}$ to $10^{-4.8}$~\smpy. The same
decreasing trends for the LMC and SMC samples occur at $\dot{M}_{gas}\sim10^{-5.3}$ and
$\dot{M}_{gas}\sim10^{-5.0}$~\smpy\ respectively and line up well with the Galactic
objects.

The trend of decreasing SiC with increasing color could result from self-absorption
within the feature due to an outer layer of cooler dust rather than a decrease in actual
SiC grain emission \citep{spe05} and may explain the observations of IRAS~04496-6958
\citep{spe06}. However, this trend also correlates with the onset of MgS emission and may
be more closely linked to the dust condensation sequence (see Sections~\ref{ldcmgs} and
5).

\subsubsection{MgS feature strength \label{ldcmgs}}

Every carbon star throughout the entire range of color-temperatures presented here
contains an apparent SiC dust feature; however, less than half of the entire sample show
a MgS band. Only six out of 29 carbon stars in the SMC sample exhibit a MgS feature, 27
out of 44 in the LMC, and 14 out of 34 for Galactic sources.

Throughout the entire sample, there appears to be an overarching
trend for redder objects to contain a stronger MgS emission relative
to the continuum. All objects with a \ctemp\ color greater than 0.9
(corresponding to a $T_{cont} \lesssim 500$ K) contain a significant
MgS dust feature while all objects with $[6.4]-[9.3]<0.6$ show no
emission from MgS, with the exception of V~Cyg, V~CrB, and T Dra in
the Milky Way. This is consistent with the condensation temperature
of MgS ($\sim$600-300~K), which is always lower than the SiC
condensation temperature.

We also find that the MgS condensation temperature depends on
pressure. Plotting the MgS strength in terms of the gas mass-loss
rate reveals that the majority of objects with an MgS feature lie
within a relatively small mass-loss range, albeit a bit offset in
terms of metallicity. About half of the Galactic spectra containing
an MgS feature have continuum dust temperatures greater than the
condensation temperature. \cite{hon04} have explained this
phenomenon in terms of an optically thin detached dust shell due to
time-variable mass loss. Such objects are identified by a blue
spectrum, low mass-loss rate, and weak MgS feature.

Each metallicity environment appears to contain a sharp transition
region where MgS emission is observed which occurs at different
mass-loss rates and color temperatures (vertical dotted lines in
Figure~\ref{fig:dust}). These offsets can be attributed to
metallicity differences and can be used to constrain the conditions
under which we would expect to observe MgS grain emission, assuming
the MgS and amorphous carbon exist in the same volume (ie. no
detached dust shell).

These transition points also correspond with the decrease in SiC emission discussed in
Section~\ref{sic}. Figures~\ref{fig:dust}a and \ref{fig:dust}b show the correlation
between SiC emission strength and the MgS feature with respect to the dust and gas
mass-loss rates. Objects containing weak MgS features mainly occupy positions near the
peaks of the SiC trend discussed above. As MgS intensity strengthens, SiC declines
suggesting that the MgS condenses in the form of a coating on top of the SiC grains, thus
suppressing the SiC emission (see Section~5).

\begin{figure*}[htp]
\begin{center}
\includegraphics[height=1.\textwidth, angle=90]{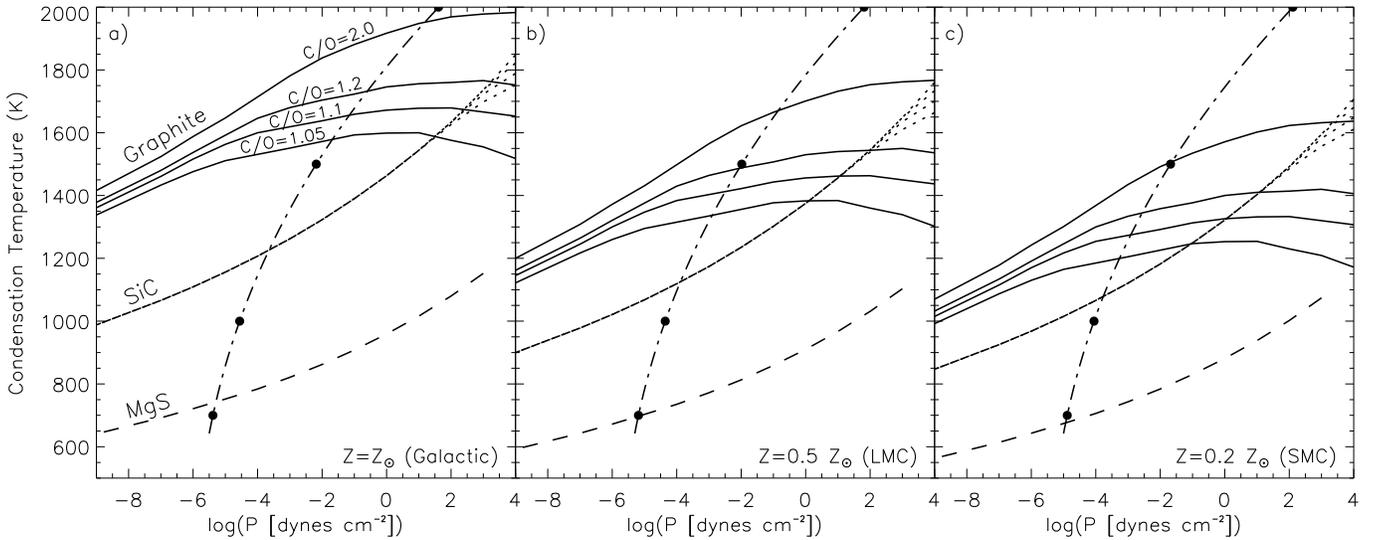}\\
\caption{{Condensation temperatures of carbon graphite, SiC, and MgS as a function of the
total pressure and C/O ratio (1.05 to 2.0) based on \cite{lod95}. Data for the LMC and
SMC are scaled from \cite{lod06}. The dash-dotted line shows a model atmosphere for a
Carbon AGB star with the following parameters: $M_*=1$~M$_{\sun}$, $L_*=10^4$~L$_{\sun}$,
$T_{\rm eff}=2400$~K, and C/O~$=1.7$ \citep{woi06}. The filled circles correspond to a
radius of 1.1~$R_*$, 1.5~$R_*$, 3~$R_*$, and 6~$R_*$ from top to bottom, respectively.
The curves in panels b and c have been shifted to compensate for metallicity effects.}}
\label{fig:tcond}
\end{center}
\end{figure*}

\section{Dust Condensation Sequence}

In order to decipher the dust condensation sequence in carbon stars, we must first
constrain various parameters in the envelope such as temperature, gas pressure (density),
C/O ratio, and abundance (metallicity). For solar metallicity stars, \cite{lod95} give
the condensation temperatures of C, SiC, and MgS for varying C/O ratio and pressure
(Figure~\ref{fig:tcond}a). At a given C/O ratio and pressure, \cite{lod06} shows that the
condensation temperatures will decrease with decreasing metallicity; however, this effect
is more pronounced for carbon than for SiC or MgS. Figures~\ref{fig:tcond}b and
\ref{fig:tcond}c plot the estimated shift in the condensation temperatures for
metallicities corresponding to the LMC and SMC, respectively. We also compare the radial
structure of a stellar model from \cite{woi06} to show the approximate regions and
sequence of various dust condensates. Furthermore, \cite{jor92} show that for a given
mass, decreasing the effective temperature or the C/O ratio will generally shift this
curve downward and to the left. A reduction in the metallicity similarly reduces the
pressure at a given radius, but the temperature structure will remain fairly constant.

Observationally, prominent MgS does not appear until the mass-loss
rates for Galactic, LMC, and SMC objects become greater than
$10^{-5.8}$, $10^{-5.4}$, and $10^{-5.1}$~\smpy, respectively
(Figure~\ref{fig:dust}d). 
We, therefore, support the conclusions from \cite{lod06} that for a
given temperature, decreasing the metallicity also leads to an
increase in the critical density at which MgS condenses.

\begin{figure}[t]
\begin{center}
\includegraphics[width=.45\textwidth]{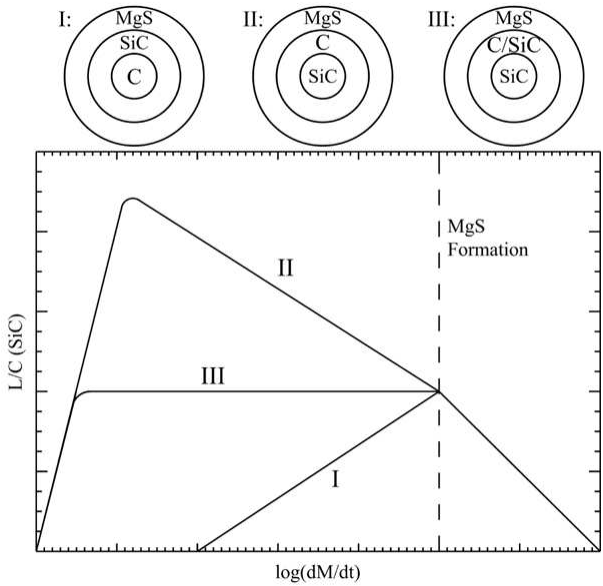}\\
\caption{{We show the three separate grain condensation sequences at the top and plot the
expected SiC-to-C ratio with mass-loss rate for each scenario. The dashed line indicates
the onset of MgS grain condensation.}} \label{fig:condseq}
\end{center}
\end{figure}

Figure~\ref{fig:condseq} is a schematic of the trends we expect to
see for three different dust condensation sequences to which this
method is sensitive. These can be differentiated by analyzing the
change in the SiC-to-C ratio with $\dot{M}_{gas}$ and taking into
account the formation of MgS (Figure~\ref{fig:dust}b):
\begin{description}
\item[I. C~$\rightarrow$~SiC~$\rightarrow$~MgS:]
In general, carbon stars will start with a low C/O ratio and a low
mass-loss rate. As the wind expands away from the star and the
temperature decreases, carbon grains condense first. As the
temperature further decreases, SiC will condense and coat the
already existing carbon grains raising the relative emission of SiC.
Finally, as mass-loss increases and the shell further expands and
cools, MgS will condense last onto existing dust grains. In this
scenario, one would expect no SiC emission for low $\dot{M}$, then a
gradually increasing SiC feature for intermediate mass-loss rates.
Higher $\dot{M}$ would be characterized by a decrease in the SiC
strength due to MgS coating the SiC dust grains.
\item[II. SiC~$\rightarrow$~C~$\rightarrow$~MgS:]
In this case, SiC would condense first at low mass-loss rates and is
characterized by a sharp increase in SiC-to-C ratio. Carbon
condenses next as mass-loss progresses and the circumstellar
envelope cools, reducing the SiC-to-C ratio. MgS forms last on the
existing existing grains further reducing the SiC strength.
\item[III. SiC~$\rightarrow$~(C, SiC)~$\rightarrow$~MgS:]
This scenario is characterized by an initially rapid increase of SiC
emission. If C/O ratio, metallicity, and density are such that SiC
and C have comparable condensation temperatures, then we would
observe a subsequent SiC-to-C ratio that is constant with $\dot{M}$
until MgS begins to condense. Afterwards, the observed SiC strength
would decline.
\end{description}

\cite{lag07} argue for a SiC~$\rightarrow$~C condensation sequence for C-stars in the
Galaxy. Figures~\ref{fig:dust}a and \ref{fig:dust}b show that the relative emission of
SiC-to-C for Galactic objects increases at low mass-loss rates; but due to the large
scatter and high uncertainties at low $\dot{M}_{gas}$, we are unable to determine with
absolute certainty whether the trend levels off or continues to increase until
$10^{-6}$~\smpy. If it does indeed flatten at $\dot{M}_{gas}\sim10^{-7.3}$~\smpy, this
would suggest a scenario where SiC forms initially; then, as the mass-loss rate
increases, C and SiC condense simultaneously. However, based on the increasing trend in
Figure~\ref{fig:dust}a as well as observational evidence of presolar grains indicating
that Galactic carbon stars produce mainly uncoated SiC grains \citep{ber87}, it is more
likely that the SiC-to-C ratio increases until $\dot{M}_{gas}\sim10^{-6}$~\smpy\
supporting a C~$\rightarrow$~SiC condensation sequence. The subsequent decline in SiC
strength correlates with an increase in MgS emission indicating that MgS is condensing
onto the SiC grain.

Due to the lack of objects with low mass-loss rates observed in the Magellanic Clouds, it
is difficult to discern whether SiC or graphite grains form initially. However, the fact
that we observe a relatively large scatter in the SiC-to-C ratio for objects without MgS
emission (especially for the SMC sample) suggests near-simultaneous condensation of SiC
and carbon, supporting a SiC~$\rightarrow$~(C, SiC) sequence. Similarly, \cite{lag07}
suggest that SiC and C form close together in the LMC, but that SiC forms last in the SMC
stars. A low initial Si abundance may also contribute to the observed low SiC strength in
these environments, particularly, in the SMC. Finally, similar to the Galactic stars, a
decline in SiC strength is observed as MgS increases for both the LMC and SMC objects.

Based on the positions of the model from \cite{woi06} in the three panels of
Figure~\ref{fig:tcond}, stars with higher metallicities would more likely follow a
C~$\rightarrow$~SiC~$\rightarrow$~MgS condensation sequence in an outflowing wind.
Lowering the photospheric temperature and/or the C/O ratios would shift the model
atmosphere curves down and to the right \citep{jor92}. This would not affect the
condensation sequence for Galactic carbon stars due to the initially large spread between
the condensation zones for the different dust species; but it could potentially flip the
C/SiC sequence shown for the Magellanic Clouds samples, especially for the SMC stars.
However, the increased efficiency of carbon dredge-up leading to relatively higher C/O
ratios at lower metallicities effectively reduces our ability to shift the LMC and SMC
models. Thus, this supports the above suggestion that carbon stars in both Magellanic
Clouds have SiC~$\rightarrow$~(C, SiC)~$\rightarrow$~MgS sequences.

\section{Conclusions \label{conclusions}}

We analyze a combined total of 107 carbon-rich AGB stars in the LMC, SMC, and Milky Way
to investigate the effects of metallicity on dust and molecular species in these objects'
circumstellar envelopes. Using modified blackbody fits, we isolated and studied the
spectral features throughout the 5-38~\um\ range including the P, Q, and R branches of
\acet\ absorption at 7.5, 13.7, and 14~\um; the 11.3~\um\ SiC emission feature; and the
30~\um\ MgS emission feature.

The modified blackbody fits deviate from the Manchester method used in SZL by providing a
more robust method of determining the continua above and below the aforementioned
spectral features. In particular, this work improves on the existing IRS studies from SZL
by allowing us to isolate the P and R branches in the 14~\um\ \acet\ band. For a given
gas mass-loss rate, we find that LMC and SMC objects have comparable \acet\ equivalent
widths, contrary to the findings of SZL. However, in agreement with SZL, we also find
that carbon stars within the LMC and SMC have much larger \acet\ equivalent widths than
Galactic carbon stars for a given $\dot{M}_{gas}$, which can be explained by a
combination of more efficient carbon dredge-up and less dust dilution at lower
metallicities.

For the 13.7~\um\ \acet\ Q branch, we show that dust dilution plays a prominent role on
the shape and strength of this feature at lower mass-loss rates. However, at higher
$\dot{M}$, \acet\ absorption overcomes the effects of dust dilution due to increasing
densities. Since veiling of the absorbing molecules by dust does not increase with the
mass-loss rates, we confirm the previous results from \cite{mat06} and SZL that the
13.7~\um\ \acet\ absorption band must originate throughout the circumstellar dust shell.
Conversely, the effects of dust veiling on the P and R branches become apparent at higher
gas mass-loss rates ($\dot{M}_{gas}\gtrsim10^{-6.5}$~\smpy). This supports the idea that
the optimal conditions for P and R branch transitions occur at depths below those
favoring Q branch transitions.

For objects with low to moderate mass-loss rates, we find that the molecular absorption
band at 14~\um\ affects the position of $\lambda_{SiC}$. The adjoining P branch of strong
acetylene absorption effectively reduces flux on the red side of the SiC feature causing
$\lambda_{SiC}$ to shift towards the blue. This effect occurs more prominently in the
Magellanic Clouds because of the increased \acet\ abundance due to enhanced carbon
dredge-up. Similarly, an unidentified absorption feature at around 10~\um\ shifts
$\lambda_{SiC}$ towards longer wavelengths for objects with $[6.4]-[9.3]\lesssim0.2$. SiC
self-absorption dominates the shift in $\lambda_{SiC}$ for objects with large dust
optical depths (\ctemp\ color) and small \acet\ absorption at 14~\um. Differences in
trends between Galactic, LMC, and SMC sources reflect the variation of \acet\ with
metallicity for a given \ctemp\ color.

Based on emission features of the SiC and MgS dust components with respect to the
mass-loss rate, we have developed a method to determine the dust condensation history for
a given metallicity system. We find evidence to support a C~$\rightarrow$~SiC
condensation sequence in Galactic objects. For C-stars within the Magellanic Clouds, we
suggest that a SiC~$\rightarrow$~(C, SiC) sequence best fits the observed data. Formation
and increase of the MgS feature with $\dot{M}_{gas}$ corresponds to a decrease in the SiC
strength indicating that MgS inhibits the SiC emission by coating the SiC grains. The
condensation of MgS in circumstellar atmosphere appears to be delayed until higher
pressures for decreasing metallicities.

The precision of much of this analysis depends on our ability to accurately determine
individual gas mass-loss rates. In this work, we make three major assumptions that lead
to uncertainties when deriving $\dot{M}_{gas}$: 1) the objects in each galaxy are
comprised of the same metallicity; 2) the dust-to-gas ratio scales with metallicity; 3)
and $\dot{M}_{gas}$ can be derived from the dust-to-gas ratio and $\dot{M}_{dust}$. Any
of these assumptions may contribute to the large scatter observed in
Figure~\ref{fig:dust}b for the non-MgS bearing objects in the Magellanic Clouds as well
as deviations from the analysis by SZL when comparing gas mass-loss rates. Therefore, we
are anticipating more accurate determinations of the gas mass-loss rate from direct
measurements of molecular species such as CO, particularly in the Magellanic Clouds.
Differences between CO-determined $\dot{M}_{gas}$ and those derived in this and previous
works will further improve our understanding of dust formation efficiencies as a function
of metallicity.

\acknowledgments

We thank Eric Lagadec and Albert Zijlstra for careful reading of the
manuscript and providing us with valuable comments. This work is
based on observations made with the \emph{Spitzer Space Telescope},
which is operated by the Jet Propulsion Laboratory, California
Institute of Technology under NASA contract 1407. Support for this
work was provided by NASA through contract number 1281150 issued by
JPL/Caltech. We are grateful to the PIs Peter Wood and Michael Egan
for making available the reduced data of their \emph{Spitzer}
observations (PIDs 3505 and 3277, respectively). We have made use of
data products from the Two Micron All Sky Survey, which is a joint
project of the University of Massachusetts and the Infrared
Processing and Analysis Center at Caltech, funded by NASA and the
National Science Foundation. This research has also made use of the
SIMBAD and VIZIER databases, operated at the Centre de Donn\'{e}es
astronomiques de Strasbourg, and the Infrared Science Archive at the
Infrared Processing and Analysis Center, which is operated by JPL.


\bibliographystyle{apj}
\bibliography{mybib}

\clearpage

\LongTables
\begin{deluxetable}{lccccccc}
\tabletypesize{\scriptsize} \tablewidth{0pc} \tablecaption{Spectroscopic Results
\label{spres}} \tablehead{\colhead{} & \colhead{\ctemp} & \colhead{EW(7.5)} &
\colhead{EW(14)} & \colhead{EW(13.7)} & \colhead{} &
\colhead{$\lambda_{SiC}$} & \colhead{}\\
\colhead{Name} & \colhead{(mag)} & \colhead{(\um)} & \colhead{(\um)} & \colhead{(\um)} &
\colhead{L/C (SiC)} & \colhead{(\um)} & \colhead{L/C (MgS)}} \startdata
\multicolumn{3}{l}{\textbf{Our Sample}}\\
IRAS 04496-6958 & 0.85$\pm$0.00 & 0.337$\pm$0.002 & 0.971$\pm$0.007 & 0.112$\pm$0.006 & ... & ... & ... \\
LI-LMC 603 & 1.52$\pm$0.01 & 0.092$\pm$0.002 & 0.152$\pm$0.004 & 0.043$\pm$0.002 & 0.042$\pm$0.001 & 11.18$\pm$0.05 & 0.426$\pm$0.002 \\
LI-LMC 1817 & 1.24$\pm$0.01 & 0.179$\pm$0.003 & 0.392$\pm$0.004 & 0.054$\pm$0.002 & 0.050$\pm$0.001 & 11.11$\pm$0.03 & 0.352$\pm$0.002 \\
MSX LMC 1205 & 0.67$\pm$0.01 & 0.123$\pm$0.002 & 0.103$\pm$0.008 & 0.066$\pm$0.005 & 0.045$\pm$0.002 & 11.19$\pm$0.07 & ... \\
MSX LMC 1209 & 0.87$\pm$0.01 & 0.259$\pm$0.006 & 0.232$\pm$0.011 & 0.074$\pm$0.007 & 0.124$\pm$0.004 & 11.12$\pm$0.04 & ... \\
MSX LMC 1213 & 0.77$\pm$0.01 & 0.127$\pm$0.003 & 0.114$\pm$0.015 & 0.043$\pm$0.013 & 0.159$\pm$0.003 & 11.09$\pm$0.03 & 0.481$\pm$0.011 \\
MSX LMC 1298 & 0.61$\pm$0.01 & 0.066$\pm$0.002 & ... & 0.050$\pm$0.003 & 0.106$\pm$0.001 & 11.39$\pm$0.02 & 0.083$\pm$0.003 \\
MSX LMC 1308 & 0.57$\pm$0.01 & 0.181$\pm$0.006 & 0.609$\pm$0.013 & 0.041$\pm$0.008 & 0.037$\pm$0.007 & 10.07$\pm$0.41 & ... \\
MSX LMC 1383 & 1.26$\pm$0.01 & 0.176$\pm$0.004 & 0.262$\pm$0.007 & 0.026$\pm$0.004 & 0.052$\pm$0.002 & 11.21$\pm$0.07 & 0.633$\pm$0.005 \\
MSX LMC 634 & 0.60$\pm$0.01 & 0.078$\pm$0.004 & 0.118$\pm$0.012 & 0.055$\pm$0.007 & 0.044$\pm$0.003 & 11.34$\pm$0.08 & ... \\
MSX LMC 768 & 0.85$\pm$0.01 & 0.122$\pm$0.002 & 0.191$\pm$0.005 & 0.042$\pm$0.003 & 0.085$\pm$0.002 & 11.27$\pm$0.03 & 0.072$\pm$0.005 \\
MSX LMC 774 & 1.08$\pm$0.01 & 0.226$\pm$0.004 & 0.273$\pm$0.005 & 0.056$\pm$0.003 & 0.075$\pm$0.002 & 11.11$\pm$0.06 & 0.349$\pm$0.006 \\
MSX LMC 783 & 0.68$\pm$0.01 & 0.114$\pm$0.003 & 0.271$\pm$0.010 & 0.048$\pm$0.008 & 0.128$\pm$0.003 & 11.13$\pm$0.04 & ... \\
MSX LMC 787 & 0.96$\pm$0.01 & 0.195$\pm$0.004 & 0.327$\pm$0.011 & 0.023$\pm$0.006 & 0.110$\pm$0.003 & 11.10$\pm$0.04 & 0.645$\pm$0.096 \\
MSX LMC 841 & 0.68$\pm$0.01 & 0.141$\pm$0.004 & 0.094$\pm$0.008 & 0.047$\pm$0.006 & 0.081$\pm$0.003 & 11.06$\pm$0.06 & ... \\
MSX LMC 937 & 0.91$\pm$0.01 & 0.130$\pm$0.002 & 0.109$\pm$0.005 & 0.058$\pm$0.003 & 0.101$\pm$0.002 & 11.23$\pm$0.03 & 0.278$\pm$0.004 \\
MSX LMC 950 & 0.88$\pm$0.01 & 0.147$\pm$0.004 & 0.361$\pm$0.009 & 0.075$\pm$0.006 & 0.112$\pm$0.004 & 11.10$\pm$0.05 & 0.287$\pm$0.017 \\
MSX LMC 971 & 1.12$\pm$0.01 & 0.166$\pm$0.005 & 0.433$\pm$0.007 & 0.040$\pm$0.005 & 0.069$\pm$0.002 & 11.19$\pm$0.04 & 0.422$\pm$0.006 \\
MSX LMC 974 & 0.74$\pm$0.01 & 0.055$\pm$0.005 & 0.223$\pm$0.009 & 0.063$\pm$0.006 & 0.049$\pm$0.003 & 11.01$\pm$0.08 & ... \\
\\
\multicolumn{3}{l}{\textbf{\cite{zij06} Sample:}}\\
GRRV 38 & 0.24$\pm$0.03 & 0.384$\pm$0.017 & 0.237$\pm$0.032 & 0.135$\pm$0.023 & 0.173$\pm$0.016 & 11.17$\pm$0.08 & ... \\
IRAS 04557-6753 & 1.07$\pm$0.01 & 0.218$\pm$0.006 & 0.195$\pm$0.007 & 0.060$\pm$0.003 & 0.091$\pm$0.004 & 11.21$\pm$0.07 & 0.461$\pm$0.004 \\
IRAS 05009-6616 & 0.90$\pm$0.00 & 0.084$\pm$0.002 & 0.111$\pm$0.004 & 0.051$\pm$0.002 & 0.122$\pm$0.001 & 11.21$\pm$0.02 & 0.267$\pm$0.003 \\
IRAS 05112-6755 & 1.12$\pm$0.01 & 0.180$\pm$0.003 & 0.386$\pm$0.004 & 0.058$\pm$0.002 & 0.052$\pm$0.001 & 11.08$\pm$0.04 & 0.317$\pm$0.002 \\
IRAS 05113-6739 & 1.05$\pm$0.02 & 0.187$\pm$0.004 & 0.085$\pm$0.007 & 0.044$\pm$0.005 & 0.066$\pm$0.002 & 11.23$\pm$0.08 & 0.299$\pm$0.004 \\
IRAS 05132-6941 & 1.05$\pm$0.01 & 0.154$\pm$0.004 & 0.219$\pm$0.010 & 0.062$\pm$0.007 & 0.094$\pm$0.006 & 11.24$\pm$0.08 & 0.138$\pm$0.006 \\
IRAS 05190-6748 & 1.25$\pm$0.01 & 0.172$\pm$0.003 & 0.251$\pm$0.006 & 0.063$\pm$0.004 & 0.044$\pm$0.005 & 11.15$\pm$0.20 & 0.418$\pm$0.001 \\
IRAS 05278-6942 & 1.11$\pm$0.01 & 0.135$\pm$0.004 & 0.295$\pm$0.009 & 0.074$\pm$0.007 & 0.067$\pm$0.005 & 11.02$\pm$0.12 & 0.502$\pm$0.002 \\
IRAS 05295-7121 & 0.95$\pm$0.01 & 0.108$\pm$0.004 & 0.139$\pm$0.007 & 0.063$\pm$0.004 & 0.066$\pm$0.003 & 11.20$\pm$0.07 & 0.169$\pm$0.003 \\
IRAS 05360-6648 & 1.06$\pm$0.01 & 0.157$\pm$0.004 & 0.279$\pm$0.007 & 0.042$\pm$0.004 & 0.125$\pm$0.003 & 11.07$\pm$0.04 & 0.691$\pm$0.005 \\
MSX LMC 219 & 1.13$\pm$0.01 & 0.256$\pm$0.003 & 0.541$\pm$0.006 & 0.048$\pm$0.004 & 0.048$\pm$0.002 & 11.18$\pm$0.05 & 0.384$\pm$0.004 \\
MSX LMC 341 & 0.96$\pm$0.01 & 0.142$\pm$0.006 & 0.166$\pm$0.007 & 0.050$\pm$0.003 & 0.124$\pm$0.002 & 11.01$\pm$0.03 & 0.672$\pm$0.008 \\
MSX LMC 349 & 0.98$\pm$0.01 & 0.113$\pm$0.003 & 0.322$\pm$0.007 & 0.045$\pm$0.003 & 0.122$\pm$0.003 & 11.04$\pm$0.04 & 0.740$\pm$0.008 \\
MSX LMC 441 & 1.12$\pm$0.01 & 0.195$\pm$0.005 & 0.407$\pm$0.007 & 0.046$\pm$0.004 & 0.082$\pm$0.002 & 11.07$\pm$0.03 & 0.647$\pm$0.004 \\
MSX LMC 443 & 0.89$\pm$0.01 & 0.261$\pm$0.004 & 0.364$\pm$0.009 & 0.037$\pm$0.006 & 0.119$\pm$0.003 & 11.11$\pm$0.03 & 0.534$\pm$0.004 \\
MSX LMC 494 & 0.55$\pm$0.01 & 0.186$\pm$0.003 & 0.065$\pm$0.004 & 0.036$\pm$0.003 & 0.095$\pm$0.002 & 11.23$\pm$0.03 & ... \\
MSX LMC 601 & 0.89$\pm$0.01 & 0.035$\pm$0.003 & 0.109$\pm$0.005 & 0.055$\pm$0.003 & 0.043$\pm$0.003 & 11.11$\pm$0.11 & ... \\
MSX LMC 679 & 0.89$\pm$0.01 & 0.161$\pm$0.003 & 0.267$\pm$0.008 & 0.035$\pm$0.005 & 0.131$\pm$0.005 & 11.07$\pm$0.07 & 0.534$\pm$0.003 \\
MSX LMC 743 & 1.03$\pm$0.01 & 0.279$\pm$0.004 & 0.425$\pm$0.007 & 0.028$\pm$0.004 & 0.107$\pm$0.003 & 11.11$\pm$0.04 & 0.701$\pm$0.003 \\
MSX LMC 749 & 0.60$\pm$0.01 & 0.146$\pm$0.003 & 0.320$\pm$0.008 & 0.067$\pm$0.005 & 0.063$\pm$0.006 & 11.02$\pm$0.15 & ... \\
MSX LMC 754 & 0.60$\pm$0.02 & 0.122$\pm$0.004 & 0.554$\pm$0.012 & 0.013$\pm$0.008 & 0.132$\pm$0.004 & 11.10$\pm$0.04 & ... \\
MSX LMC 967 & 0.59$\pm$0.01 & 0.055$\pm$0.004 & 0.168$\pm$0.012 & 0.062$\pm$0.011 & 0.065$\pm$0.006 & 11.23$\pm$0.21 & ... \\
NGC 1978 IR1 & 0.55$\pm$0.01 & 0.122$\pm$0.002 & 0.383$\pm$0.007 & 0.057$\pm$0.003 & 0.088$\pm$0.002 & 11.08$\pm$0.05 & ... \\
TRM 72 & 0.78$\pm$0.01 & 0.051$\pm$0.004 & 0.121$\pm$0.007 & 0.069$\pm$0.004 & 0.064$\pm$0.002 & 11.18$\pm$0.05 & ... \\
TRM 88 & 0.76$\pm$0.02 & 0.121$\pm$0.006 & 0.178$\pm$0.006 & 0.055$\pm$0.003 & 0.058$\pm$0.007 & 10.97$\pm$0.13 & ... \\
\\
\multicolumn{3}{l}{\textbf{\cite{slo06} Sample:}}\\
MSX SMC 033 & 0.51$\pm$0.01 & 0.052$\pm$0.002 & 0.081$\pm$0.006 & 0.050$\pm$0.003 & 0.059$\pm$0.002 & 10.96$\pm$0.06 & ... \\
MSX SMC 036 & 0.76$\pm$0.01 & 0.299$\pm$0.005 & 0.401$\pm$0.008 & 0.054$\pm$0.005 & 0.042$\pm$0.002 & 11.06$\pm$0.04 & ... \\
MSX SMC 044 & 0.53$\pm$0.01 & 0.041$\pm$0.003 & ... & 0.066$\pm$0.008 & 0.079$\pm$0.003 & 10.98$\pm$0.07 & ... \\
MSX SMC 054 & 0.75$\pm$0.01 & 0.236$\pm$0.002 & 0.209$\pm$0.006 & 0.057$\pm$0.004 & 0.139$\pm$0.001 & 11.19$\pm$0.01 & 0.177$\pm$0.020 \\
MSX SMC 060 & 0.96$\pm$0.00 & 0.215$\pm$0.003 & ... & 0.037$\pm$0.003 & 0.030$\pm$0.002 & 11.27$\pm$0.08 & 0.228$\pm$0.027 \\
MSX SMC 062 & 0.65$\pm$0.02 & 0.129$\pm$0.006 & 0.055$\pm$0.012 & 0.053$\pm$0.007 & 0.055$\pm$0.003 & 11.13$\pm$0.09 & ... \\
MSX SMC 066 & 0.43$\pm$0.01 & 0.091$\pm$0.002 & 0.464$\pm$0.012 & 0.040$\pm$0.004 & 0.025$\pm$0.002 & 11.24$\pm$0.04 & ... \\
MSX SMC 091 & 0.65$\pm$0.02 & 0.358$\pm$0.010 & 0.329$\pm$0.022 & 0.059$\pm$0.017 & 0.121$\pm$0.007 & 11.23$\pm$0.07 & ... \\
MSX SMC 093 & 0.46$\pm$0.01 & 0.241$\pm$0.004 & 1.077$\pm$0.014 & 0.027$\pm$0.008 & 0.016$\pm$0.003 & 10.81$\pm$0.11 & ... \\
MSX SMC 105 & 0.87$\pm$0.01 & 0.297$\pm$0.004 & 0.085$\pm$0.006 & 0.029$\pm$0.004 & 0.120$\pm$0.003 & 11.12$\pm$0.04 & 0.487$\pm$0.014 \\
MSX SMC 142 & 0.29$\pm$0.03 & 0.165$\pm$0.007 & ... & 0.090$\pm$0.030 & 0.131$\pm$0.016 & 11.56$\pm$0.28 & ... \\
MSX SMC 159 & 0.87$\pm$0.01 & 0.204$\pm$0.002 & 0.285$\pm$0.005 & 0.047$\pm$0.004 & 0.124$\pm$0.001 & 11.04$\pm$0.02 & 0.417$\pm$0.010 \\
MSX SMC 162 & 0.50$\pm$0.01 & 0.091$\pm$0.004 & 0.393$\pm$0.016 & 0.006$\pm$0.011 & 0.078$\pm$0.004 & 11.09$\pm$0.10 & ... \\
MSX SMC 163 & 0.68$\pm$0.01 & 0.171$\pm$0.003 & 0.276$\pm$0.006 & 0.054$\pm$0.003 & 0.119$\pm$0.002 & 11.15$\pm$0.02 & 0.236$\pm$0.008 \\
MSX SMC 198 & 0.62$\pm$0.01 & 0.141$\pm$0.003 & 0.009$\pm$0.011 & 0.073$\pm$0.005 & 0.055$\pm$0.002 & 11.10$\pm$0.05 & ... \\
MSX SMC 200 & 0.42$\pm$0.01 & 0.007$\pm$0.003 & 0.095$\pm$0.007 & 0.069$\pm$0.005 & ... & ... & ... \\
MSX SMC 202 & 0.45$\pm$0.02 & 0.158$\pm$0.006 & 0.078$\pm$0.019 & 0.002$\pm$0.017 & 0.037$\pm$0.007 & 10.65$\pm$0.23 & ... \\
MSX SMC 209 & 0.70$\pm$0.01 & 0.198$\pm$0.002 & 0.350$\pm$0.007 & 0.047$\pm$0.003 & ... & ... & ... \\
MSX SMC 232 & 0.63$\pm$0.01 & 0.100$\pm$0.002 & 0.270$\pm$0.009 & 0.061$\pm$0.004 & 0.009$\pm$0.002 & 10.92$\pm$0.15 & ... \\
\\
\multicolumn{3}{l}{\textbf{\cite{lag07} Sample:}}\\
GM780 & 0.41$\pm$0.01 & 0.340$\pm$0.010 & 0.164$\pm$0.024 & 0.035$\pm$0.016 & 0.246$\pm$0.007 & 11.07$\pm$0.05 & ... \\
IRAS 00554-7351 & 0.94$\pm$0.01 & 0.145$\pm$0.005 & 0.186$\pm$0.009 & 0.078$\pm$0.007 & 0.085$\pm$0.005 & 11.27$\pm$0.09 & 0.291$\pm$0.004 \\
ISO 00548 & 0.41$\pm$0.01 & 0.060$\pm$0.003 & 0.023$\pm$0.008 & 0.064$\pm$0.005 & 0.086$\pm$0.002 & 11.30$\pm$0.03 & ... \\
ISO 00573 & 0.30$\pm$0.01 & 0.167$\pm$0.006 & 0.197$\pm$0.013 & 0.078$\pm$0.008 & 0.007$\pm$0.007 & 11.72$\pm$0.56 & ... \\
ISO 01019 & 0.41$\pm$0.02 & 0.137$\pm$0.004 & 0.001$\pm$0.012 & 0.039$\pm$0.009 & 0.056$\pm$0.005 & 11.25$\pm$0.10 & ... \\
LEGC 105 & 0.36$\pm$0.02 & 0.094$\pm$0.002 & 0.198$\pm$0.010 & 0.054$\pm$0.005 & 0.044$\pm$0.003 & 11.09$\pm$0.07 & ... \\
NGC 419 IR1 & 0.49$\pm$0.01 & 0.152$\pm$0.002 & 0.766$\pm$0.008 & 0.041$\pm$0.004 & ... & ... & ... \\
NGC 419 LE 16 & 0.34$\pm$0.02 & 0.090$\pm$0.004 & 0.185$\pm$0.016 & 0.066$\pm$0.012 & 0.071$\pm$0.004 & 11.23$\pm$0.06 & ... \\
NGC 419 MIR1 & 1.36$\pm$0.01 & 0.278$\pm$0.005 & 0.500$\pm$0.007 & 0.048$\pm$0.004 & 0.023$\pm$0.003 & 11.40$\pm$0.13 & 0.409$\pm$0.003 \\
RAW 960 & 0.23$\pm$0.01 & 0.226$\pm$0.004 & 0.141$\pm$0.026 & 0.064$\pm$0.009 & 0.063$\pm$0.005 & 11.17$\pm$0.09 & ... \\
\\
\multicolumn{3}{l}{\textbf{\cite{slo03} Sample:}}\\
AFGL 2392 & 0.58$\pm$0.00 & 0.081$\pm$0.001 & 0.028$\pm$0.002 & 0.022$\pm$0.001 & 0.312$\pm$0.000 & 11.19$\pm$0.00 & 0.336$\pm$0.002 \\
AFGL 3099 & 1.21$\pm$0.00 & 0.063$\pm$0.001 & 0.008$\pm$0.001 & 0.067$\pm$0.001 & 0.130$\pm$0.000 & 11.43$\pm$0.00 & 0.427$\pm$0.001 \\
AFGL 341 & 1.67$\pm$0.01 & 0.080$\pm$0.001 & 0.042$\pm$0.001 & 0.050$\pm$0.001 & 0.032$\pm$0.000 & 12.12$\pm$0.01 & 0.318$\pm$0.001 \\
AFGL 940 & 0.77$\pm$0.00 & 0.040$\pm$0.001 & 0.010$\pm$0.003 & 0.041$\pm$0.002 & 0.176$\pm$0.001 & 11.27$\pm$0.01 & 0.311$\pm$0.002 \\
C 2178 & 0.42$\pm$0.00 & 0.155$\pm$0.001 & 0.158$\pm$0.004 & 0.041$\pm$0.002 & 0.200$\pm$0.001 & 11.23$\pm$0.01 & ... \\
C 2429 & 0.19$\pm$0.00 & 0.166$\pm$0.001 & ... & 0.032$\pm$0.002 & 0.614$\pm$0.001 & 11.36$\pm$0.00 & ... \\
Case 181 & 0.61$\pm$0.00 & 0.047$\pm$0.001 & 0.081$\pm$0.001 & 0.052$\pm$0.001 & 0.199$\pm$0.000 & 11.25$\pm$0.00 & 0.186$\pm$0.000 \\
DO 16793 & 1.29$\pm$0.00 & 0.043$\pm$0.000 & 0.252$\pm$0.001 & 0.048$\pm$0.001 & 0.078$\pm$0.000 & 11.30$\pm$0.01 & 0.454$\pm$0.000 \\
DO 40123 & 0.29$\pm$0.00 & 0.311$\pm$0.001 & 0.101$\pm$0.002 & 0.057$\pm$0.001 & 0.275$\pm$0.000 & 11.08$\pm$0.00 & ... \\
HV Cas & 0.35$\pm$0.00 & 0.146$\pm$0.000 & 0.004$\pm$0.002 & 0.043$\pm$0.001 & 0.216$\pm$0.000 & 11.34$\pm$0.00 & ... \\
IRAS 21489+5301 & 1.36$\pm$0.01 & 0.102$\pm$0.001 & 0.072$\pm$0.002 & 0.044$\pm$0.002 & 0.018$\pm$0.001 & 12.09$\pm$0.02 & 0.347$\pm$0.001 \\
IRC+00365 & 0.67$\pm$0.00 & 0.077$\pm$0.000 & ... & 0.032$\pm$0.001 & 0.178$\pm$0.000 & 11.32$\pm$0.00 & 0.206$\pm$0.000 \\
IRC+20326 & 0.89$\pm$0.00 & 0.009$\pm$0.001 & 0.030$\pm$0.001 & 0.003$\pm$0.000 & 0.078$\pm$0.000 & 11.44$\pm$0.01 & 0.354$\pm$0.000 \\
IRC+40540 & 1.16$\pm$0.00 & 0.064$\pm$0.000 & 0.001$\pm$0.000 & 0.036$\pm$0.000 & 0.148$\pm$0.000 & 11.33$\pm$0.00 & 0.310$\pm$0.000 \\
IRC+50096 & 0.68$\pm$0.00 & 0.111$\pm$0.001 & 0.246$\pm$0.001 & 0.031$\pm$0.001 & 0.257$\pm$0.000 & 11.13$\pm$0.00 & 0.144$\pm$0.000 \\
IRC-10095 & 0.16$\pm$0.00 & 0.292$\pm$0.001 & ... & 0.013$\pm$0.002 & 0.140$\pm$0.001 & 11.35$\pm$0.01 & ... \\
IRC-10122 & 0.38$\pm$0.00 & 0.093$\pm$0.000 & 0.140$\pm$0.003 & 0.031$\pm$0.001 & 0.243$\pm$0.000 & 11.15$\pm$0.00 & ... \\
RU Vir & 0.51$\pm$0.00 & 0.076$\pm$0.001 & 0.085$\pm$0.001 & 0.011$\pm$0.001 & 0.320$\pm$0.000 & 11.08$\pm$0.00 & ... \\
RY Dra & 0.11$\pm$0.00 & 0.115$\pm$0.000 & ... & 0.081$\pm$0.001 & 0.099$\pm$0.000 & 11.61$\pm$0.00 & ... \\
R For & 0.36$\pm$0.00 & 0.142$\pm$0.001 & 0.073$\pm$0.002 & 0.023$\pm$0.001 & 0.264$\pm$0.000 & 11.13$\pm$0.00 & ... \\
R Scl & 0.27$\pm$0.00 & 0.390$\pm$0.001 & 0.712$\pm$0.003 & 0.077$\pm$0.001 & 0.154$\pm$0.001 & 11.09$\pm$0.00 & ... \\
SS Vir & 0.36$\pm$0.00 & 0.420$\pm$0.000 & 0.274$\pm$0.009 & 0.058$\pm$0.004 & 0.317$\pm$0.001 & 11.17$\pm$0.00 & ... \\
S Cep & 0.30$\pm$0.00 & 0.160$\pm$0.000 & ... & 0.045$\pm$0.001 & 0.253$\pm$0.001 & 11.15$\pm$0.01 & ... \\
S Sct & 0.14$\pm$0.00 & 0.293$\pm$0.001 & 0.022$\pm$0.002 & 0.098$\pm$0.002 & 0.160$\pm$0.000 & 11.15$\pm$0.01 & ... \\
TU Tau & 0.08$\pm$0.00 & ... & ... & 0.128$\pm$0.007 & 0.270$\pm$0.001 & 11.44$\pm$0.01 & ... \\
T Dra & 0.51$\pm$0.00 & 0.162$\pm$0.001 & 0.137$\pm$0.003 & 0.026$\pm$0.001 & 0.344$\pm$0.000 & 11.07$\pm$0.00 & 0.149$\pm$0.002 \\
U Cam & 0.11$\pm$0.00 & 0.065$\pm$0.000 & 0.095$\pm$0.002 & 0.075$\pm$0.001 & 0.279$\pm$0.000 & 11.34$\pm$0.00 & ... \\
V460 Cyg & 0.16$\pm$0.00 & 0.125$\pm$0.000 & 0.067$\pm$0.004 & 0.077$\pm$0.003 & 0.049$\pm$0.001 & 11.56$\pm$0.02 & ... \\
VX And & 0.12$\pm$0.00 & 0.214$\pm$0.000 & 0.260$\pm$0.004 & 0.101$\pm$0.002 & 0.115$\pm$0.000 & 11.39$\pm$0.00 & ... \\
V Aql & 0.11$\pm$0.00 & 0.146$\pm$0.001 & 0.156$\pm$0.002 & 0.097$\pm$0.001 & 0.098$\pm$0.000 & 11.38$\pm$0.00 & ... \\
V CrB & 0.40$\pm$0.00 & 0.166$\pm$0.001 & 0.180$\pm$0.003 & 0.053$\pm$0.001 & 0.387$\pm$0.000 & 11.12$\pm$0.00 & 0.140$\pm$0.005 \\
V Cyg & 0.52$\pm$0.00 & 0.196$\pm$0.000 & 0.351$\pm$0.003 & 0.036$\pm$0.001 & 0.309$\pm$0.000 & 11.01$\pm$0.00 & 0.060$\pm$0.001 \\
W Ori & 0.06$\pm$0.00 & 0.058$\pm$0.000 & 0.006$\pm$0.001 & 0.082$\pm$0.001 & 0.144$\pm$0.000 & 11.47$\pm$0.00 & ... \\
Y CVn & 0.15$\pm$0.00 & 0.179$\pm$0.001 & 0.029$\pm$0.002 & 0.067$\pm$0.002 & 0.068$\pm$0.000 & 11.50$\pm$0.01 & ... \\
\enddata
\tablecomments{For all objects in each sample, we show the \ctemp\ color as well as
strengths of the molecular and dust features. The molecular features are measured in
terms of equivalent widths (EW) in microns while the dust emission bands show a
line-to-continuum ratio (L/C). We also give the central wavelength of the SiC feature
($\lambda_{SiC}$).}
\end{deluxetable}

\clearpage \LongTables
\begin{deluxetable}{lrrcc|lrrcc}
\tabletypesize{\scriptsize} \tablewidth{0pc} \tablecaption{Mass-loss Rates
\label{tbl:mloss}} \tablehead{\colhead{Name} &
\colhead{$\log\dot{M}_{dust}$\tablenotemark{a}} &
\colhead{$\log\dot{M}_{gas}$\tablenotemark{b}} & \colhead{$\dot{M}_{dust}$} &
\colhead{Type\tablenotemark{c}} & \colhead{Name} &
\colhead{$\log\dot{M}_{dust}$\tablenotemark{a}} &
\colhead{$\log\dot{M}_{gas}$\tablenotemark{b}} & \colhead{$\dot{M}_{dust}$} & \colhead{Type\tablenotemark{c}}\\
\colhead{} & \multicolumn{2}{c}{(\smpy)} & \colhead{Ref.} & \colhead{(Ref.)} & \colhead{}
& \multicolumn{2}{c}{(\smpy)} & \colhead{Ref.} & \colhead{(Ref.)}} \startdata
\multicolumn{3}{l}{\textbf{Our Sample}}\\
IRAS 04496-6958 & -7.55 & -5.25 & 1 & ... & MSX LMC 768 & -7.55 & -5.25 & 1 & ... \\
LI-LMC 603 & -6.76 & -4.46 & 1 & ... & MSX LMC 774 & -7.25 & -4.95 & 1 & ... \\
LI-LMC 1817 & -7.06 & -4.76 & 1 & ... & MSX LMC 783 & -7.80 & -5.50 & 1 & ... \\
MSX LMC 1205 & -7.81 & -5.51 & 1 & ... & MSX LMC 787 & -7.40 & -5.10 & 1 & ... \\
MSX LMC 1209 & -7.52 & -5.22 & 1 & ... & MSX LMC 841 & -7.80 & -5.50 & 1 & ... \\
MSX LMC 1213 & -7.67 & -5.37 & 1 & ... & MSX LMC 937 & -7.47 & -5.17 & 1 & ... \\
MSX LMC 1298 & -7.91 & -5.61 & 1 & ... & MSX LMC 950 & -7.51 & -5.21 & 1 & ... \\
MSX LMC 1308 & -7.98 & -5.68 & 1 & ... & MSX LMC 971 & -7.20 & -4.90 & 1 & ... \\
MSX LMC 1383 & -7.04 & -4.74 & 1 & ... & MSX LMC 974 & -7.71 & -5.41 & 1 & ... \\
MSX LMC 634 & -7.93 & -5.62 & 1 & ... &  &  &  &  & \\
\\
\multicolumn{3}{l}{\textbf{\cite{zij06} Sample:}}\\
GRRV 38 & -8.64 & -6.34 & 1 & R (2) & MSX LMC 441 & -7.26 & -4.96 & 2 & R (2) \\
IRAS 04557-6753 & -7.22 & -4.92 & 2 & N (2) & MSX LMC 443 & -7.60 & -5.30 & 2 & R (2) \\
IRAS 05009-6616 & -7.33 & -5.03 & 2 & N (2) & MSX LMC 494 & -8.15 & -5.85 & 2 & R (2) \\
IRAS 05112-6755 & -7.05 & -4.74 & 2 & N (2) & MSX LMC 601 & -7.59 & -5.29 & 2 & R (2) \\
IRAS 05113-6739 & -7.22 & -4.92 & 2 & N (2) & MSX LMC 679 & -7.54 & -5.24 & 2 & R (2) \\
IRAS 05132-6941 & -7.32 & -5.02 & 2 & R (2) & MSX LMC 743 & -7.43 & -5.13 & 2 & R (2) \\
IRAS 05190-6748 & -6.94 & -4.64 & 2 & N (2) & MSX LMC 749 & -7.70 & -5.40 & 2 & R (2) \\
IRAS 05278-6942 & -6.90 & -4.60 & 2 & R (2) & MSX LMC 754 & -8.07 & -5.77 & 2 & R (2) \\
IRAS 05295-7121 & -7.35 & -5.05 & 2 & N (2) & MSX LMC 967 & -7.80 & -5.49 & 2 & R (2) \\
IRAS 05360-6648 & -7.39 & -5.09 & 2 & N (2) & NGC 1978 IR1 & -7.94 & -5.64 & 2 & N (2) \\
MSX LMC 219 & -7.30 & -5.00 & 2 & R (2) & TRM 72 & -7.54 & -5.24 & 2 & N (2) \\
MSX LMC 341 & -7.43 & -5.13 & 2 & R (2) & TRM 88 & -7.78 & -5.48 & 2 & R (2) \\
MSX LMC 349 & -7.40 & -5.10 & 2 & R (2) &  &  &  &  &  \\
\\
\multicolumn{3}{l}{\textbf{\cite{slo06} Sample:}}\\
MSX SMC 033 & -7.65 & -4.95 & 2 & R (2) & MSX SMC 142 & -8.57 & -5.87 & 2 & R (2) \\
MSX SMC 036 & -7.76 & -5.06 & 2 & N (2) & MSX SMC 159 & -7.62 & -4.92 & 2 & N (2) \\
MSX SMC 044 & -7.70 & -5.00 & 2 & R (2) & MSX SMC 162 & -8.19 & -5.49 & 2 & N (2) \\
MSX SMC 054 & -7.74 & -5.05 & 2 & ... & MSX SMC 163 & -7.78 & -5.08 & 2 & R (2) \\
MSX SMC 060 & -7.55 & -4.85 & 2 & N (2) & MSX SMC 198 & -8.12 & -5.43 & 2 & R (2) \\
MSX SMC 062 & -7.80 & -5.10 & 2 & R (2) & MSX SMC 200 & -7.89 & -5.19 & 2 & ... \\
MSX SMC 066 & -7.77 & -5.07 & 2 & N (2) & MSX SMC 202 & -8.36 & -5.66 & 2 & ... \\
MSX SMC 091 & -8.12 & -5.43 & 2 & ... & MSX SMC 209 & -7.74 & -5.05 & 2 & R (2) \\
MSX SMC 093 & -8.60 & -5.90 & 2 & ... & MSX SMC 232 & -7.84 & -5.14 & 2 & N (2) \\
MSX SMC 105 & -7.67 & -4.97 & 2 & R (2) &  &  &  &  \\
\\
\multicolumn{3}{l}{\textbf{\cite{lag07} Sample:}}\\
GM780 & -8.22 & -5.52 & 2 & R (2) & LEGC 105 & -8.26 & -5.56 & 2 & R (2) \\
IRAS 00554-7351 & -7.19 & -4.49 & 2 & N (2) & NGC 419 IR1 & -8.00 & -5.30 & 2 & N (2) \\
ISO 00548 & -7.96 & -5.26 & 2 & R (2) & NGC 419 LE 16 & -8.59 & -5.89 & 2 & N (2) \\
ISO 00573 & -8.51 & -5.81 & 2 & R (2) & NGC 419 MIR1 & -7.07 & -4.37 & 2 & R (2) \\
ISO 01019 & -8.26 & -5.56 & 2 & R (2) & RAW 960 & -8.76 & -6.06 & 2 & R (2) \\
\\
\multicolumn{3}{l}{\textbf{\cite{slo03} Sample:}}\\
AFGL 2392 & -7.74 & -5.74 & 3 & M (5) & RU Vir & -7.96 & -5.96 & 4 & M (7) \\
AFGL 3099 & -7.10 & -5.10 & 3 & M (6) & RY Dra & -9.82 & -7.82 & 4 & S (7) \\
AFGL 341 & -7.02 & -5.02 & 4 & M (6) & R For & -8.10 & -6.10 & 3 & M (7) \\
AFGL 940 & -7.49 & -5.49 & 4 & ... & R Scl & -9.15 & -7.15 & 3 & S (7) \\
C 2178 & -7.95 & -5.95 & 4 & ... & SS Vir & -8.74 & -6.74 & 4 & S (7) \\
C 2429 & -9.11 & -7.11 & 4 & I (7) & S Cep & -8.71 & -6.71 & 4 & M (7) \\
Case 181 & -7.32 & -5.32 & 4 & I (6) & S Sct & -10.08 & -8.08 & 4 & S (7) \\
DO 16793 & -12.06 & -10.06 & 4 & ... & TU Tau & -9.85 & -7.85 & 4 & S (7) \\
DO 40123 & -8.58 & -6.58 & 4 & M (6) & T Dra & -8.41 & -6.41 & 4 & M (7) \\
HV Cas & -8.05 & -6.05 & 4 & M (5) & U Cam & -9.75 & -7.75 & 4 & S (7) \\
IRAS 21489+5301 & -6.80 & -4.80 & 4 & ... & V460 Cyg & -10.51 & -8.51 & 4 & S (7) \\
IRC+00365 & -7.50 & -5.50 & 4 & M (6) & VX And & -9.33 & -7.33 & 4 & S (7) \\
IRC+20326 & -7.27 & -5.27 & 4 & ... & V Aql & -9.35 & -7.35 & 4 & S (7) \\
IRC+40540 & -7.08 & -5.08 & 4 & M (7) & V CrB & -9.08 & -7.08 & 4 & M (7) \\
IRC+50096 & -7.81 & -5.81 & 4 & M (7) & V Cyg & -8.16 & -6.16 & 4 & M (7) \\
IRC-10095 & -8.52 & -6.52 & 4 & ... & W Ori & -9.87 & -7.87 & 4 & S (7) \\
IRC-10122 & -7.92 & -5.92 & 4 & M (6) & Y CVn & -9.92 & -7.92 & 4 & S (7) \\
\enddata

\tablenotetext{a}{Dust mass-loss rates are either results of dust radiative transfer
models of a given source or taken from relations based off of these models.}

\tablenotetext{b}{Gas mass-loss rates were derived using Equations~\ref{mdotlmc} through
\ref{mdotmw}.}

\tablenotetext{c}{Objects in the LMC and SMC exhibit Mira-like pulsations: R stand for
\emph{regular} pulsation and N stands for \emph{non-regular} pulsation. For Galactic
objects, M stand for \emph{Mira}, I stand for \emph{Irregular}, and S stand for
\emph{Semiregular}.}

\tablerefs{(1) Relation from \cite{van07}; (2) \cite{gro07}; (3) \cite{leb97}; (4)
Relation from \cite{leb97}; (5) \cite{gro02}; (6) \cite{lou93}; (7) \cite{ber05}}

\end{deluxetable}

\end{document}